\documentstyle[prb,twocolumn,aps,epsbox]{revtex}
\input{epsf.tex}
\newcommand{\bm}[1]{\mbox{\boldmath$#1$}}
\begin{document} \draft 
\twocolumn[
\title{Tricritical Behavior in the Extended Hubbard Chains}
\author{Masaaki Nakamura\cite{address}}
\address{Institute for Solid State Physics,
University of Tokyo, Roppongi, Tokyo 106-8666, Japan}
\date{Phys. Rev. B {\bf 61}, 16377 (2000)}\maketitle
\begin{abstract}
\widetext\leftskip=0.10753\textwidth \rightskip\leftskip
 Phase diagrams of the one-dimensional extended Hubbard model (including
 nearest-neighbor interaction $V$) at half- and quarter-filling are
 studied by observing level crossings of excitation spectra using the
 exact diagonalization.  This method is based on the Tomonaga-Luttinger
 liquid theory including logarithmic corrections which stem from the
 renormalization of the Umklapp- and the backward-scattering effects.
 Using this approach, the phase boundaries are determined with high
 accuracy, and then the structure of the phase diagram is clarified. At
 half-filling, the phase diagram consists of two
 Berezinskii-Kosterlitz-Thouless (BKT) transition lines and one Gaussian
 transition line in the charge sector, and one spin-gap transition line. 
 This structure reflects the U(1) $\otimes$ SU(2) symmetry of the
 electron system. Near the $U=2V$ line, the Gaussian and the spin-gap
 transitions take place independently from the weak- to the
 intermediate-coupling region, but these two transition lines are
 coupled in the strong-coupling region. This result demonstrates
 existence of a tricritical point and a bond-charge-density-wave (BCDW)
 phase between charge- and spin-density-wave (CDW, SDW) phases. To
 clarify this mechanism of the transition, we also investigate effect of
 a correlated hopping term which plays a role to enlarge BCDW and
 bond-spin-density-wave (BSDW) phases.  At quarter-filling, a similar
 crossover phenomenon also takes place in the large-$V$ region involving
 spin-gap and BKT-type metal-insulator transitions.
\end{abstract}
] \narrowtext
\section{INTRODUCTION}

One-dimensional (1D) electron systems have been extensively studied
motivated not only by theoretical interest but also by the discovery of
quasi-1D conductors and high-$T_{\rm c}$ superconductivity.  In the 1D
electron systems, due to the charge-spin separation, the low-energy
excitations in the charge and the spin sectors may have gaps
independently, and then various phases can appear.  However, phenomena
caused by interplay between these two degrees of freedom have not been
fully understood even in simple models.  In this paper, we turn our
attention to the phase transitions in the so-called extended Hubbard
model (EHM),
\begin{eqnarray}
 {\cal H}_{\rm EHM}&=&
 -t\sum_{is}(c^{\dag}_{is} c_{i+1,s}+\mbox{H.c.})\nonumber\\
 &&+U\sum_i n_{i\uparrow}n_{i\downarrow}+V\sum_i n_{i}n_{i+1},
\label{eqn:tUV}
\end{eqnarray}
at half- and quarter-filling, where both charge and spin gaps can open.

The EHM at half-filling has been studied using various approaches. In
the weak-coupling limit, the phase diagram is analytically obtained by
the g-ology\cite{Emery,Solyom,Voit,Voit92} (see Appendix
\ref{sec:weak-coupling}). According to the result, there appear
insulating charge- (CDW) and spin-density-wave (SDW) phases, and
metallic phases where the singlet superconducting (SS) or the triplet
superconducting (TS) correlation is dominant [see
Fig.~\ref{fig:g-ology}(b)].  On the other hand, in the strong-coupling
limit, the perturbation theory gives the phase boundary of the CDW-SDW
transition\cite{Bari,Hirsch,Dongen} and of the phase
separation.\cite{Emery76,Fowler,Lin-H} The rest of region has been
discussed by numerical
analysis.\cite{Hirsch,Fourcade-S,Cannon-F,Cannon-S-F,Lin-G-C-F-G,GPZhang}
However, the phase diagrams are not fully understood, because the charge
and the spin gaps open exponentially slow [see Eqs.~(\ref{eqn:spin-gap})
and (\ref{eqn:charge-gap})], which makes it difficult to determine the
phase boundaries by the conventional finite-size scaling method.
Especially for the transition between the CDW and the SDW phases, even
the property of the transition itself is not clear, because the
transition is of the second order in the weak-coupling theory, while it
is of the first order in the strong-coupling theory.

\begin{figure}[b]
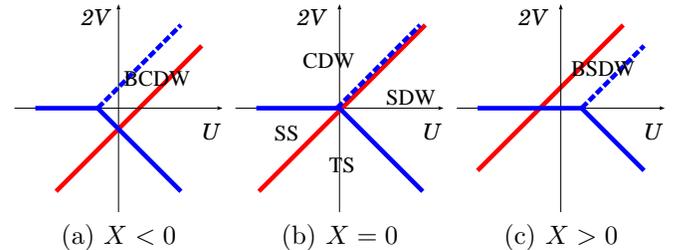

\noindent
\begin{tabular}{ccc}
\epsfxsize=2.8cm \leavevmode \epsfbox{pdch1.eps}&
\epsfxsize=2.8cm \leavevmode \epsfbox{pdch0.eps}&
\epsfxsize=2.8cm \leavevmode \epsfbox{pdch2.eps}\\
 (a) $X<0$ & (b) $X=0$ & (c) $X>0$
\end{tabular}
 \caption{Weak-coupling phase diagrams of the EHM including the
 correlated hopping term (Ref.~\protect{\ref{Japaridze-K}}). The phase
 diagrams are given by combinations of a Y-shaped structure for the
 charge part [two BKT lines (solid) and one Gaussian line (dashed)] and
 an I-shaped one for the spin part.}  \label{fig:g-ology}
\end{figure}

Recently, the author has clarified the mechanism of the CDW-SDW
transition.\cite{Nakamura_99} According to the result, the phase
boundary consists of two independent transition lines, and the crossover
of the CDW-SDW transition is related with whether these two transition
lines are separated or coupled.  The result also demonstrates the
existence of the bond-charge-density-wave (BCDW)\cite{BOW_comment} phase
in the very narrow region between the CDW and the SDW states.  In this
paper, we not only give the details of the letter, but also clarify the
entire phase diagram of the EHM at half-filling.

In order to clarify the above scenario for the phase transition between
the CDW and the SDW phases, we also consider generalizing the EHM by
adding the following correlated hopping interactions:
\cite{Campbell-G-L,Simon-A,Japaridze-K,Aligia-H-B-O}
\begin{equation}
 {\cal H}_{X}=X\sum_{is}
 (c^{\dag}_{is} c_{i+1,s}+\mbox{H.c.})(n_{i,-s}-n_{i+1,-s})^2.
\label{eqn:X-term}
\end{equation}
This interaction can be derived as a site-off-diagonal element of the
Coulomb integral.\cite{Campbell-G-L} Especially, the three-body part is
justified as an effective interaction of the three-band
model.\cite{Simon-A} The weak-coupling phase diagram is known by the
g-ology\cite{Japaridze-K} as shown in Figs.~\ref{fig:g-ology} (a) and
(c) (see Appendix \ref{sec:weak-coupling}). In these phase diagrams, the
two transition lines between the CDW and the SDW phases do not
synchronize. And a BCDW or a bond-spin-density-wave\cite{Japaridze}
(BSDW) phase appears. The analysis of the generalized model will clarify
the tricritical behavior in the pure EHM. The final results are shown in
Fig.~\ref{fig:PD1}.

A charge-gap phase is known to appear not only at half-filling but also
at quarter-filling,\cite{Mila-Z,Penc-M,Sano-O,Clay-S-C} due to the
effect of the Umklapp scattering in the higher
order.\cite{Giamarchi,Kolomeisky-S,Oshikawa-Y-A,Yamanaka-O-A} In this
case, the interplay between charge and spin instabilities is also
expected. In fact, we will conclude that a crossover phenomenon also
exists in the large-$V$ region at quarter-filling (see
Fig.~\ref{fig:PD2}).

Throughout this paper, we use the level-crossing approach to determine
the phase boundaries.\cite{Nakamura_99,Okamoto-N,Nomura-O,Nomura95,%
Nakamura-N-K,Nakamura_98,Nakamura-K-N,Bursill-M-H} This method is based
on the Tomonaga-Luttinger (TL) liquid theory\cite{Haldane} (which is
equivalent to the $c=1$ conformal field theory) including the
logarithmic corrections, which stem from the renormalization of the
Umklapp- and the backward-scattering effects.  In the theoretical
scheme, the transition points are identified by the level crossing of
the excitation spectra in the finite-size ring with size dependence
${\cal O}(L^{-2})$, where $L$ is the system size.  Therefore, the phase
boundaries are obtained with high accuracy, using the numerical data of
finite-size clusters.

This paper is organized as follows. In Sec.~\ref{sec:theory}, we review
the level-crossing approach based on the TL liquid theory and the
renormalization group, developed in
Refs.~\ref{Okamoto-N}-\ref{Nakamura-K-N}.  In Sec.~\ref{sec:SYMMETRY},
we discuss the discrete symmetries of wave functions to connect
excitation spectra and the corresponding physical states.  In
Sec.~\ref{sec:PHASES}, we discuss the character of the phases that
appear in the phase diagrams.  In Sec.~\ref{sec:HALF}, we analyze the
instabilities of the EHM at half-filling, and clarify the phase diagram.
In Sec.~\ref{sec:QUARTER}, we analyze the metal-insulator transition at
quarter-filling. Finally, a summary and discussions are given in
Sec.~\ref{sec:SUMMARY}.  In Appendix \ref{sec:weak-coupling}, we briefly
explain the traditional g-ology analysis for the generalized EHM at
half-filling.

\section{PHASE BOUNDARIES}\label{sec:theory}

First, let us perform a general argument for 1D electron systems based
on the bosonization
theory.\cite{Emery,Solyom,Voit,Voit92,Schulz,Senechal} The continuous
fermion fields are defined by $c_{js}/\sqrt{a}\rightarrow \psi_{{\rm
L},s}(x)+\psi_{{\rm R},s}(x)$ (the lattice constant $a\rightarrow 0$,
$x=ja$) with
\begin{equation}
 \psi_{r,s}(x)=\frac{U_{r,s}}{\sqrt{2\pi\alpha}}
  {\rm e}^{{\rm i} r k_{\rm F} x}{\rm e}^{{\rm i}/\sqrt{2}\cdot
 [r(\phi_{\rho}+s\phi_{\sigma})-\theta_{\rho}-s\theta_{\sigma}]},
  \label{eqn:fermion_op}
\end{equation}
where $r={\rm R},{\rm L}$ and $s=\uparrow, \downarrow$ refer to $+$ and
$-$ in that order. $\alpha$ is a short-distance cutoff.  $k_{\rm F}$ is
the Fermi wave number defined by $k_{\rm F}\equiv\pi n/2$, with $n$
being the electron density.  The field $\phi_{\nu}$ and the dual field
$\theta_{\nu}$ of the charge ($\nu=\rho$) and the spin ($\nu=\sigma$)
degrees of freedom satisfy the relation
\begin{equation}
[\phi_{\mu}(x),\theta_{\nu}(x')]=
 -\frac{{\rm i}\pi}{2}\delta_{\mu\nu}{\rm sign}(x-x').
\end{equation}
$U_{r,s}$ ensures anticommutation relations of the different fermion
fields.\cite{Haldane,Senechal} These operators are Hermitian and satisfy
the relation
\begin{equation}
\{U_{r,s},U_{r',s'}\}=2\delta_{r,r'}\delta_{s,s'}.
 \label{eqn:ladder_op}
\end{equation}
Using the formalism, the low-energy behavior of 1D electron system can
be described by the sine-Gordon models for the charge and the spin
sectors. When $2q$ ($q$: integer) electrons contribute to the Umklapp
scattering, the effective Hamiltonian for the system with length $L$ is
given by
\begin{eqnarray}
   {\cal H}&=&\sum_{\nu=\rho,\sigma}
   \frac{v_{\nu}}{2\pi}\int_0^L {\rm d}x
  \left[K_{\nu}(\partial_x \theta_{\nu})^2
       +K_{\nu}^{-1}(\partial_x \phi_{\nu})^2\right]\nonumber\\
 &+&\frac{2 g_{1\perp}}{(2\pi\alpha)^2}
  \int_0^L {\rm d}x \cos[\sqrt{8}\phi_{\sigma}(x)]\nonumber\\
 &+&\frac{2 g_{3\perp}}{(2\pi\alpha)^2}
  \int_0^L {\rm d}x \cos[q\sqrt{8}\phi_{\rho}(x)+\delta x]\nonumber\\
 &+&\frac{2 g_{3\parallel}}{(2\pi \alpha)^2}
 \int_0^L {\rm d}x
 \cos[q\sqrt{8}\phi_{\rho}(x)+\delta x]\cos[\sqrt{8}\phi_{\sigma}(x)],
  \label{eqn:eff_Ham}
\end{eqnarray}
where $v_{\nu}$ and $K_{\nu}$ are the velocity and the Gaussian
coupling, respectively for each sector.  $g_{1\perp}$ and $g_{3\perp}$
denote the amplitude of the backward and the Umklapp scattering,
respectively.  The Umklapp term vanishes except for the case
$\delta\equiv 2p\pi-4qk_{\rm F}=0$ where $p$ is also an integer, and
$p/q$ is an irreducible fraction.  Thus, the electron filling that a
charge gap can open is quantized to commensurate cases $n=p/q$.
\cite{Giamarchi,Kolomeisky-S,Oshikawa-Y-A} This condition can also be
derived from the generalized Lieb-Schultz-Mattis theorem.
\cite{Oshikawa-Y-A,Yamanaka-O-A} In this paper, we consider $q=1$
(half-filling) and $q=2$ (quarter-filling) cases.  At half-filling, in
the weak-coupling limit, the couplings of the backward and the Umklapp
scattering for the EHM are identified as $g_{1\perp}=-g_{3\perp}=U-2V$
(see Appendix \ref{sec:weak-coupling}).

In addition to the $g_{3\perp}$ term, there exists another Umklapp
operator with coupling constant $g_{3\parallel}$, which transfers finite
spin.\cite{Voit92,Cannon-F,Kolomeisky-S} In the weak-coupling limit,
this parameter is identified as $g_{3\parallel}=-2V$. In the present
analysis, we will not consider this term explicitly, because the scaling
dimension of this term is always higher than that of the other nonlinear
terms in Eq.~(\ref{eqn:eff_Ham}).  However, in the strong-coupling
region, a charge-spin coupling effect may appear due to this term.

If these nonlinear terms are absent
($g_{1\perp}=g_{3\perp}=g_{3\parallel}=0$), the excitation spectra and
their wave numbers in the finite-size system are described by
\begin{eqnarray}
 E-E_0&=&
 \frac{2\pi v_{\rho}}{L}x_{\rho}+\frac{2\pi v_{\sigma}}{L}x_{\sigma},
 \label{eqn:energy}\\
 P-P_0&=&\frac{2\pi}{L}(s_{\rho}+s_{\sigma})+2m_{\rho}k_{\rm F},
\end{eqnarray} 
where the scaling dimensions and the conformal spins are given by
\begin{eqnarray}
 x_{\nu}&=&
  \frac{1}{2}\left(\frac{n_{\nu}^2}{K_{\nu}}+m_{\nu}^2K_{\nu}\right)
  +N_{\nu}+\bar{N}_{\nu},\label{eqn:scaling_dim}\\
 s_{\nu}&=&n_{\nu}m_{\nu}+N_{\nu}-\bar{N}_{\nu}.
\end{eqnarray}
Here $n_{\rho}$ is the change of $2n_{\rho}$ electrons, and $n_{\sigma}$
is the total $z$-spins $S^z_T=n_{\sigma}$.  $m_{\rho}$ ($m_{\sigma}$)
denotes the number of particles moved from the left charge (spin) Fermi
point to the right one.  The non-negative integers $N_{\nu}$ and
$\bar{N}_{\nu}$ are the particle-hole excitations near the right and the
left Fermi points, respectively.  The scaling dimensions are related to
the critical exponents for the correlation functions in the large
distance as
\begin{equation}
\langle{\cal O}_i(r){\cal O}_i(r')\rangle
\propto|r-r'|^{-2(x_{\rho i}+x_{\sigma i})},
\label{eqn:correlation_func}
\end{equation}
where the operator is given by
\begin{eqnarray}
{\cal O}_{i}&\equiv&
{\cal O}_{n_{\rho},m_{\rho}}^{\rho}{\cal O}_{n_{\sigma},m_{\sigma}}^{\sigma}
\nonumber\\
{\cal O}_{n_{\nu},m_{\nu}}^{\nu}
&\equiv&{\rm e}^{{\rm i}\sqrt{2}(n_{\nu}\theta_{\nu}+m_{\nu}\phi_{\nu})},
\label{eqn:generall_op}
\end{eqnarray}
or linear combinations of these operators. Therefore, there are
one-to-one correspondences between the excitation spectra and the
operators.

Now we turn our attention to the excitation spectra which correspond to
the following operators:
\begin{mathletters}
\begin{eqnarray}
{\cal O}_{\nu 0}&\equiv&
-{\textstyle\frac{4}{K_{\nu}}}\bar{\partial}\phi_{\nu}\partial\phi_{\nu},
\label{eqn:marginal}\\
{\cal O}_{\nu 1}&\equiv&\sqrt{2}\cos(q\sqrt{2}\phi_{\nu})
 \propto{\cal O}_{0,q}^{\nu}+{\cal O}_{0,-q}^{\nu},
\label{eqn:cos}\\
{\cal O}_{\nu 2}&\equiv&\sqrt{2}\sin(q\sqrt{2}\phi_{\nu})
 \propto{\cal O}_{0,q}^{\nu}-{\cal O}_{0,-q}^{\nu},
\label{eqn:sin}\\
{\cal O}_{\nu 3}&\equiv&\exp({\rm i}\sqrt{2}\theta_{\nu})
 ={\cal O}_{1,0}^{\nu},
\label{eqn:exp}
\end{eqnarray}
\end{mathletters}
where ${\cal O}_{\nu 0}$ is the ``marginal field'',\cite{Kadanoff-B} and
the derivatives are defined by
$\partial,\bar{\partial}\equiv(v_{\nu}^{-1}\partial_{\tau}\mp{\rm
i}\partial_x)/2$ with imaginary time $\tau$. This operator corresponds
to particle-hole excitations near the right and the left Fermi points
$(N_{\nu}=\bar{N}_{\nu}=1)$.  ${\cal O}_{\nu 1}$ and ${\cal O}_{\nu 2}$
are linear combinations of current excitations ($m_{\nu}=\pm q$). ${\cal
O}_{\nu 3}$ is an excitation accompanying variation of the number of
electrons or spins ($n_{\nu}=\pm 1$). We have to choose antiperiodic
boundary conditions $\psi_{r,s}(x+L)=-\psi_{r,s}(x)$ to extract the
excitation spectra for ${\cal O}_{\nu 1}$ and ${\cal O}_{\nu 2}$ fields,
when $q$ is odd, and ${\cal O}_{\nu 3}$
field,\cite{Nakamura-N-K,Nakamura_98,Nakamura-K-N} because the phase
fields satisfy the following boundary conditions:\cite{Haldane}
\begin{mathletters}
\begin{eqnarray}
 \phi_{\nu}(x+L)&=&\phi_{\nu}(x)-\sqrt{2}\pi n_{\nu},\\
 \theta_{\nu}(x+L)&=&\theta_{\nu}(x)+\sqrt{2}\pi m_{\nu},
\end{eqnarray}
\end{mathletters}
and the Fermi operator is given by these phase fields as in
Eq.~(\ref{eqn:fermion_op}).

The effects of the $g_{1\perp}$ and the $g_{3\perp}$ terms in
Eq.~(\ref{eqn:eff_Ham}) are renormalized in the scaling dimensions
$x_{\nu}$ as logarithmic corrections which are analyzed by the
renormalization group (RG) equations derived under the change of the
cutoff $\alpha\rightarrow {\rm e}^{{\rm d}l}\alpha$
(Ref.~\ref{Kosterlitz}). Within the one-loop order, the RG equations are
given by
\begin{mathletters}\label{eqn:RGE}
\begin{eqnarray}
 \frac{{\rm d}y_{\nu 0}(l)}{{\rm d}l}&=&-y_{\nu\phi}^{\ 2}(l),\\
 \frac{{\rm d}y_{\nu \phi}(l)}{{\rm d}l}&=&-y_{\nu 0}(l) y_{\nu\phi}(l),
  \label{eqn:RGE_Gaussian}
\end{eqnarray}
\end{mathletters} 
where $y_{\rho 0}(0)=2(q^2K_{\rho}-1)$, $y_{\sigma
0}(0)=2(K_{\sigma}-1)$, $y_{\rho\phi}(0)=g_{3\perp}/\pi v_{\rho}$,
$y_{\sigma\phi}(0)=g_{1\perp}/\pi v_{\sigma}$, and we have set $l=\ln
L$. These equations determine the RG flow diagrams.  Note that there is
a difference between the cases for the charge and the spin sectors
reflecting their symmetries (see Fig.~\ref{fig:RGflow}).  In the
following subsections, we discuss the phase transitions for each sector
described by these RG flow diagrams.

\subsection{Spin-gap transition}\label{sec:spin-gap}

\begin{figure}[h]
\noindent
\begin{center}
\epsfxsize=2.5in \leavevmode \epsfbox{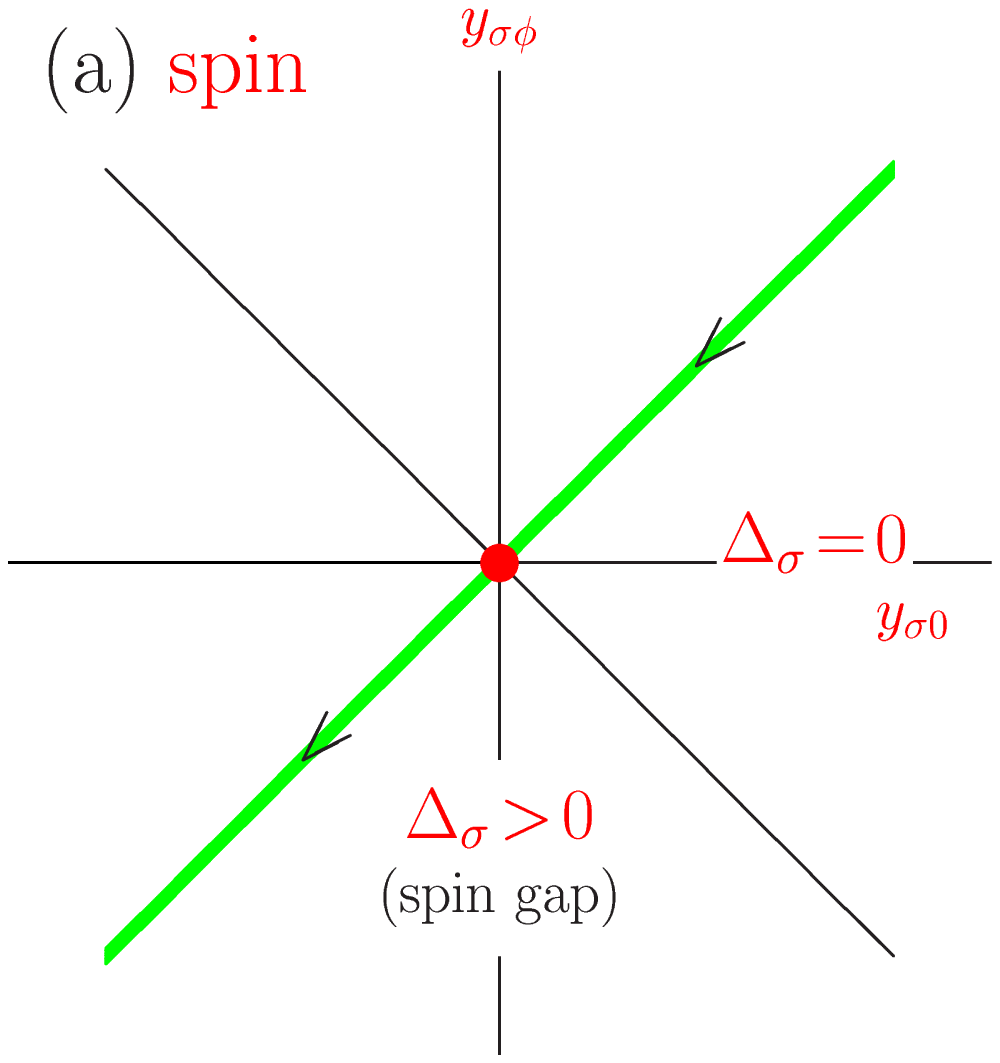}\\
 \vspace{0.5cm}
\epsfxsize=2.5in \leavevmode \epsfbox{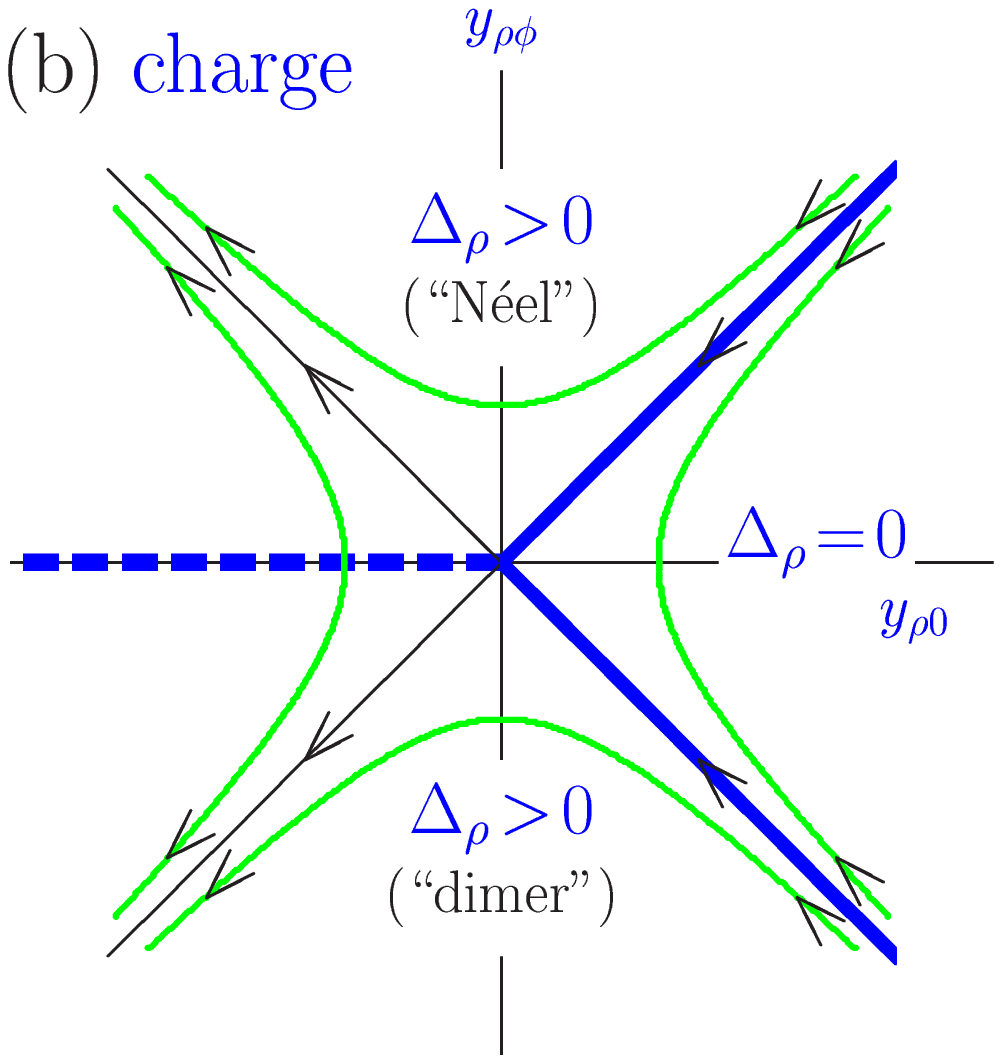}
\end{center}
\caption{RG flow diagram for (a) the spin and (b) the charge degrees of
 freedom.  For the spin sector, the RG flow is fixed on the $y_{\sigma
 0}=y_{\sigma\phi}$ line due to the SU(2) symmetry, and the spin-gap
 transition takes place at $y_{\sigma 0}= y_{\sigma\phi}=0$.  For the
 charge sector, BKT-type transitions take place on the $y_{\rho 0}=\pm
 y_{\rho\phi}$ lines with $y_{\rho 0}>0$. When $q=1$, the $y_{\rho
 0}=+y_{\rho\phi}$ line corresponds to the SU(2) symmetry of the
 $\eta$-paring. The $y_{\rho 0}=-y_{\rho\phi}$ line reflects the hidden
 symmetry of the sine-Gordon model.  A Gaussian transition occurs on the
 $y_{\rho\phi}=0$ line with $y_{\rho 0}<0$.}  \label{fig:RGflow}
\end{figure}

First, we consider the phase transition in the spin degree of freedom
($\nu=\sigma$).  The spin sector with an SU(2) symmetry belongs to the
universality class of the level-$1$ SU(2) Wess-Zumino-Novikov-Witten
(WZNW) model.\cite{Tsvelik} In this case, the RG flow in
Fig.~\ref{fig:RGflow} (a) is fixed on the $y_{\sigma
0}(l)=y_{\sigma\phi}(l)$ line. Then for $y_{\sigma 0}(l)>0$, the
exponent is renormalized as $K_{\sigma}^{*}=1$, and the solution of
Eq.~(\ref{eqn:RGE}) is obtained as
\begin{equation}
y_{\sigma 0}(l)=\frac{y_{\sigma 0}(0)}{y_{\sigma 0}(0) l+1},
 \label{eqn:renormalized_coupling}
\end{equation}
where $y_{\sigma 0}(0)$ is the bare coupling constant.  Combining the
renormalized coupling and the operator-product-expansion coefficients,
the singlet ($x_{\sigma 1}$) and the triplet ($x_{\sigma 2,3}$)
excitation spectra split as\cite{Affleck-G-S-Z,Giamarchi-S,Okamoto-N}
\begin{mathletters}
\begin{eqnarray}
 x_{\sigma 1}(l)&=&\frac{1}{2}+\frac{3}{4}y_{\sigma 0}(l),\\
 x_{\sigma 2,3}(l)&=&\frac{1}{2}-\frac{1}{4}y_{\sigma 0}(l).
\end{eqnarray}
 \label{eqn:sclngdim_spin}
\end{mathletters}
When $y_{\sigma 0}(0)<0$, $y_{\sigma 0}(l)$ is renormalized as
$y_{\sigma 0}(l\rightarrow\infty)=-\infty$, then a spin gap appears. At
the critical point [$y_{\sigma 0}(0)=0$], there are no logarithmic
corrections in the excitation spectra.  Therefore, the critical point is
obtained by the intersection of the singlet and the triplet excitation
spectra ($x_{\sigma 1}=x_{\sigma
2,3}$).\cite{Okamoto-N,Nakamura-N-K,Nakamura_98,Nakamura-K-N} This level
crossing corresponds to the condition for the spin-gap phase boundary
$g_{1\perp}=g_{\sigma}=0$ in the standard g-ology analysis, (see
Appendix \ref{sec:weak-coupling} and Table \ref{tbl:gology_vs_lc}).

The asymptotic behavior of the spin gap against a parameter of a model
$\lambda$ near the critical point $\lambda_{\rm c}$ is obtained by the
two-loop RG equation and the definition of correlation length $y_{\sigma
0}(\ln\xi)\sim-1$ as\cite{Nakamura-N-K,Nakamura-K-N}
\begin{equation}
 \Delta_{\sigma}\sim v_{\sigma}/\xi\propto
  \sqrt{\lambda-\lambda_{\rm c}}
  \exp[-\mbox{const}/(\lambda-\lambda_{\rm c})],
  \label{eqn:spin-gap}
\end{equation}
where we have used a relation $\lambda-\lambda_{\rm c}\propto |y_{\sigma
0}(0)|$.  Note that this is the same behavior as that of the spin gap in
the 1D negative-$U$ Hubbard model at half-filling.\cite{Ovchinnikov}

\subsection{Berezinskii-Kosterlitz-Thouless transition}
\subsubsection{SU(2) symmetric case}\label{sec:charge_BKT_SU2}
Next, we consider the instabilities in the charge sector ($\nu=\rho$),
which are described by the RG flow diagram given in
Fig.~\ref{fig:RGflow}(b).  At half-filling ($q=1$) and $V=0$ (the
Hubbard model), the sign of the on-site interaction $U$ in the
Hamiltonian (\ref{eqn:tUV}) is reversed by the following canonical
transformation:\cite{Shiba}
\begin{equation}
c_{j\uparrow}\rightarrow c_{j\uparrow},\ \ \ 
c_{j\downarrow}\rightarrow (-1)^j c_{j\downarrow}^{\dag}.
\end{equation}
This transformation also projects the spin ($\zeta$-pairing) operators
onto the $\eta$-pairing ones
\begin{equation}
  \eta_i^{+}=(-1)^i c_{i\uparrow}^{\dag}c_{i\downarrow}^{\dag},\ \
  \eta_i^{-}=(-1)^i c_{i\downarrow}c_{i\uparrow},\ \
  \eta_i^{z}={\textstyle \frac{1}{2}}(n_i-1),
  \label{eqn:eta-pairing}
\end{equation}
without losing the SU(2) symmetry.\cite{Yang-Z} It follows from
Eq.~(\ref{eqn:fermion_op}) this transformation corresponds to the
replacement of the indices as $\rho\leftrightarrow\sigma$.  Therefore,
the spin part of the sine-Gordon model of Eq.~(\ref{eqn:eff_Ham}) is
mapped onto the charge part, and the operators ${\cal O}_{\rho 1}$ and
${\cal O}_{\rho 2}$, ${\cal O}_{\rho 3}$ denote the ``singlet'' and the
``triplet'' for the charge part, respectively.  Thus, the exponent is
renormalized as $K_{\rho}^{*}=1$ for $U<0$, and the charge gap opens for
$U>0$.

In the case when the SU(2) symmetry in the charge sector is broken by
finite $V$, ${\cal O}_{\rho 1}$, ${\cal O}_{\rho 2}$, and ${\cal
O}_{\rho 3}$ refer to ``dimer'', ``N\'{e}el'', and ``doublet'',
respectively, by following Ref.~\ref{Nomura-O}. Then, if the initial
value of the RG flow moves across the SU(2)-symmetric line
[$y_{\rho\phi}(0)=y_{\rho 0}(0)>0$], a Berezinskii-Kosterlitz-Thouless
(BKT)-type transition\cite{Berezinskii,Kosterlitz-T,Kosterlitz} occurs
between the TL liquid phase and the twofold-degenerate gapped state.
For this transition, one can show that a charge gap opens
as\cite{Kosterlitz}
\begin{equation}
 \Delta_{\rho}\propto\exp(-{\rm const}/\sqrt{\lambda-\lambda_{\rm c}}),
  \label{eqn:charge-gap}
\end{equation}
where $|\lambda-\lambda_{\rm c}|\propto t$, and $t\equiv
|y_{\rho\phi}(l)|/y_{\rho 0}(l)-1$ stands for the deviation from the BKT
critical line.  Note that Eq.~(\ref{eqn:charge-gap}) is a different
asymptotic behavior from that of the spin-gap transition described by
Eq.~(\ref{eqn:spin-gap}), so that we discriminate the spin-gap
transition from BKT-type transitions in this paper.  The critical point
for this BKT transition can be obtained without calculations, because it
is fixed by the SU(2) symmetry of the Hamiltonian. In the case of the
EHM, the BKT transition line is fixed on the $V=0$ line for $U<0$.

Now, we consider the region for $y_{\rho\phi}(l)<0$.  The sine-Gordon
model has a symmetry under the transformation to reverse the sign of the
nonlinear term $\cos\sqrt{8}\phi_{\rho}$.  This transformation
corresponds to the shift of the phase fields:
\begin{equation}
 \phi_{\rho}\rightarrow\phi_{\rho}+\pi/\sqrt{8}.
\label{eqn:sign_change}
\end{equation}
This operation interchanges the roles of the operators ${\cal O}_{\rho
1}$ and ${\cal O}_{\rho 2}$ as
\begin{mathletters}\label{eqn:opes_change}
\begin{eqnarray}
\cos\sqrt{2}\phi_{\rho}&\rightarrow&-\sin\sqrt{2}\phi_{\rho},\\
\sin\sqrt{2}\phi_{\rho}&\rightarrow&\cos\sqrt{2}\phi_{\rho}.
\end{eqnarray}
\end{mathletters}
Therefore, this symmetry indicates that the SU(2)-symmetric line in the
RG flow diagram is mapped onto the opposite side of the $y_{\rho 0}$
axis, and another BKT transition may occur at $y_{\rho\phi}(0)=-y_{\rho
0}(0)<0$. We call the symmetry of this BKT line ``hidden SU(2)
symmetry''.  Since this symmetry originates from that of the sine-Gordon
model, it is not contained explicitly in the original Hamiltonian.  The
renormalized scaling dimensions of ${\cal O}_{\rho 1}$, ${\cal O}_{\rho
2}$, and ${\cal O}_{\rho 3}$ near the critical line of the hidden SU(2)
symmetry are calculated as follows:\cite{Giamarchi-S,Nomura-O}
\begin{mathletters}
\begin{eqnarray}
 x_{\rho 1}(l)&=&\frac{1}{2}-\frac{1}{4}y_{\rho 0}(l)(1+2t),\\
 x_{\rho 2}(l)&=&\frac{1}{2}
  +\frac{3}{4}y_{\rho 0}(l)(1+{\textstyle\frac{2}{3}}t),\\
 x_{\rho 3}(l)&=&\frac{1}{2}-\frac{1}{4}y_{\rho 0}(l).
\end{eqnarray}
 \label{eqn:sclngdim_charge_BKT}
\end{mathletters}
Therefore, the critical point for the hidden SU(2) BKT transition can be
determined by the level crossing between the ``dimer'' and the
``doublet'' excitation spectra ($x_{\rho 1}=x_{\rho 3}<x_{\rho
2}$).\cite{Nomura-O} This level crossing corresponds to the condition
$g_{3\perp}=-g_{\rho}<0$ in the g-ology analysis (see Appendix
\ref{sec:weak-coupling} and Table \ref{tbl:gology_vs_lc}).

\subsubsection{Non-SU(2) symmetric case}\label{sec:non-su2}

We consider the BKT transition for $q\ge 2$ case. This situation may
appear in the metal-insulator transition at quarter-filling ($q=2$).  In
this case, the critical line no longer has an SU(2) symmetry.  However,
by replacing the variables as $\phi_{\rho}'=q\phi_{\rho}$,
$\theta_{\rho}'=\theta_{\rho}/q$, and $K_{\rho}'=q^2 K_{\rho}$, the
sine-Gordon model for the charge part of Eq.~(\ref{eqn:eff_Ham}) is
mapped onto the case of $q=1$. Then, the BKT transition between the TL
liquid and the $2q$-fold-degenerate gapped state takes place when the
renormalized exponent becomes $K_{\rho}^{*}=1/q^2$.  The scaling
dimensions for the ${\cal O}_{\rho 1}$ and the ${\cal O}_{\rho 2}$
fields near the BKT critical line remain unchanged, while the ${\cal
O}_{\rho 3}$ field changes as
\begin{equation}
 x_{\rho 3}(l)=q^2\left[\frac{1}{2}-\frac{1}{4}y_{\rho 0}(l)\right].
\end{equation}
Therefore, the BKT critical point corresponding to $y_{\rho\phi}=y_{\rho
0}>0$ is given by the level-crossing point of $x_{\rho 2}=x_{\rho
3}/q^2<x_{\rho 1}$.

There is another excitation spectrum that can be used to determine the
BKT critical point. This is the ``marginal field'' (\ref{eqn:marginal})
whose renormalized scaling dimension is given by\cite{Nomura95}
\begin{equation}
 x_{\rho 0}(l)=2-y_{\rho 0}(l)\left(1+{\textstyle\frac{4}{3}}t\right).
\end{equation}
In this case, the critical point can be determined by the level crossing
of $x_{\rho 0}=4x_{\rho 2}$.

\subsection{Gaussian transition}

In addition to these BKT-type transitions, a Gaussian transition occurs
at $y_{\rho\phi}(0)=0$ and $y_{\rho 0}(0)<0$. This is a second-order
transition between the two gapped states which corresponds to the
different fixed points [$y_{\rho\phi}(l\rightarrow\infty)=\pm\infty$],
and the gap vanishes just on the critical point. The transition point is
given by the level crossing between the ``dimer'' and the ``N\'{e}el''
excitations ($x_{\rho 1}=x_{\rho 2}<x_{\rho 3}$), because the ${\cal
O}_{\rho 1}$ and the ${\cal O}_{\rho 2}$ fields interchange their roles
at $y_{\rho\phi}(0)=0$ as was explained in
Sec.~\ref{sec:charge_BKT_SU2}.  In the g-ology, this level crossing
corresponds to the condition $g_{3\perp}=0$ with $g_{\rho}<0$ (see
Appendix \ref{sec:weak-coupling} and Table \ref{tbl:gology_vs_lc}).
Since the nonlinear term vanishes on the critical line, there is no
effect of the renormalization.  Therefore, the scaling dimensions on the
Gaussian line are given by
\begin{mathletters}
\begin{eqnarray}
 x_{\rho 1}&=&x_{\rho 2}=\frac{K_{\rho}}{2},\\
 x_{\rho 3}&=&\frac{1}{2K_{\rho}},
\end{eqnarray}
 \label{eqn:sclngdim_charge_gaussian}
\end{mathletters}
without logarithmic corrections, and $K_{\rho}<1$ is satisfied.  The
asymptotic form of the gap near the Gaussian transition point can be
obtained by solving Eq.~(\ref{eqn:RGE_Gaussian}) with an approximation
$y_{\rho 0}(l)\approx y_{\rho 0}(0)$ and definition of the correlation
length $|y_{\rho\phi}(\ln\xi)|\sim 1$ as\cite{Nomura-O}
\begin{equation}
 \Delta_{\rho}\sim v_{\rho}/\xi
  \propto |\lambda-\lambda_{\rm c}|^{\frac{1}{2(1-K_{\rho})}},
\label{eqn:gaussian_trans}
\end{equation}
where we have used the relation
$y_{\rho\phi}(0)\propto\lambda-\lambda_{\rm c}$.  The valley of the gap
becomes steeper as $K_{\rho}$ is decreased.

\begin{table}
\begin{tabular}{c|c|c}
          & g-ology           & level crossing\\\hline
 spin gap
& $g_{1\perp}(=g_{\sigma})=0$ & $x_{\sigma 1}=x_{\sigma 2,3}$\\\hline
 SU(2)BKT
& $g_{3\perp}=g_{\rho}>0$  & $x_{\rho 2}=x_{\rho 3}<x_{\rho 1}$\\
 hidden SU(2)BKT
& $g_{3\perp}=-g_{\rho}<0$   & $x_{\rho 1}=x_{\rho 3}<x_{\rho 2}$\\
 Gaussian
& $g_{3\perp}=0,g_{\rho}<0$ & $x_{\rho 1}=x_{\rho 2}<x_{\rho 3}$
\end{tabular}
\caption{Correspondence between the g-ology and the level-crossing
 approach at half-filling. The scaling dimensions $x_{\nu i}$ correspond
 to excitation spectra of singlet($x_{\sigma 1}$), triplet($x_{\sigma
 2,3}$), ``dimer''($x_{\rho 1}$), ``N\'{e}el''($x_{\rho 2}$), and
 ``doublet''($x_{\rho 3}$) states. Examples of these level crossings are
 shown in Figs. \protect{\ref{fig:check_sg}} and
 \protect{\ref{fig:LEVEL_CROSS}}.}
 \label{tbl:gology_vs_lc}
\end{table}

\section{Discrete Symmetries}\label{sec:SYMMETRY}

To perform the level-crossing analysis discussed in
Sec.~\ref{sec:theory}, we need to identify the relevant excitation
spectra. For this purpose, we discuss the discrete symmetries of wave
functions corresponding to the excited states. The physical meaning of
these symmetries will be clarified in Sec.~\ref{sec:PHASES}.

The discrete symmetries are defined under particle-hole (${\cal C}:
c_{is}\rightarrow(-1)^ic^{\dag}_{is}$), space-inversion (${\cal P}:
c_{is}\rightarrow c_{L-i+1,s}$), and spin-reversal (${\cal T}:
c_{is}\rightarrow c_{i,-s}$) transformations.  They give eigenvalues
$\pm 1$.  In addition, shift operation by one site (${\cal S}:
c_{is}\rightarrow c_{i+1,s}$) is defined which has an eigenvalue ${\rm
e}^{{\rm i}k}$. The symmetries of wave functions can be explained by
combining those of the ground state and those of the operators for the
excited states.\cite{Nakamura-K-N} For the ground state of the EHM, we
choose periodic (antiperiodic) boundary conditions when $N/2$ is odd
(even), where $N$ is the number of electrons.  Then, according to the
Perron-Frobenious theorem, the discrete symmetries of the ground state
are ${\cal C}={\cal P}={\cal T}=1$ and $k=0$, if we choose the
representation for the basis and use the symmetry operations defined in
Ref.~\ref{Nakamura-K-N}.

\begin{table}
\begin{tabular}{c|c|rrrrr}
 &operators&${\cal C}$& ${\cal P}$& ${\cal T}$& $k$& BC\\ \hline G.S.  &
 $1$ & $ 1$& $ 1$ & $ 1$ & $0$ & $\pm 1$\\
 marginal &
 $-\frac{4}{K_{\nu}}\bar{\partial}\phi_{\nu}\partial\phi_{\nu}$ & $ 1$&
 $ 1$ & $ 1$ & $0$ & $\pm 1$\\
 CDW &
 $\sin\sqrt{2}\phi_{\rho}\cdot\cos\sqrt{2}\phi_{\sigma}$ &
 $-1$& $-1$ & $ 1$ & $2k_{\rm F}$ & $\pm 1$\\
 SDW$_{z}$ &
 $\cos\sqrt{2}\phi_{\rho}\cdot\sin\sqrt{2}\phi_{\sigma}$ &
 $-1$& $-1$ & $-1$ & $2k_{\rm F}$ & $\pm 1$\\
 SDW$_{\pm}$ &
 $\cos\sqrt{2}\phi_{\rho}\cdot\exp\pm{\rm i}\sqrt{2}\theta_{\sigma}$ &
 $ *$& $ 1$ & $ *$ & $2k_{\rm F}$ & $\pm 1$\\
 BCDW &
 $\cos\sqrt{2}\phi_{\rho}\cdot\cos\sqrt{2}\phi_{\sigma}$ &
 $ 1$& $ 1$ & $ 1$ & $2k_{\rm F}$ & $\pm 1$\\
 BSDW$_{z}$ &
 $\sin\sqrt{2}\phi_{\rho}\cdot\sin\sqrt{2}\phi_{\sigma}$ &
 $ 1$& $ 1$ & $-1$ & $2k_{\rm F}$ & $\pm 1$\\
 BSDW$_{\pm}$ &
 $\sin\sqrt{2}\phi_{\rho}\cdot\exp\pm{\rm i}\sqrt{2}\theta_{\sigma}$ &
 $ *$& $-1$ & $ *$ & $2k_{\rm F}$ & $\pm 1$\\
 SS &
 $\exp{\rm i}\sqrt{2}\theta_{\rho}\cdot\cos\sqrt{2}\phi_{\sigma}$ &
 $ *$& $ 1$ & $ 1$ & $0$ & $\pm 1$\\
 TS$_0$ &
 $\exp{\rm i}\sqrt{2}\theta_{\rho}\cdot\sin\sqrt{2}\phi_{\sigma}$ &
 $ *$& $-1$ & $-1$ & $0$ & $\pm 1$\\
 \ \ TS$_{\pm1}$ &
 $\exp{\rm i}\sqrt{2}\theta_{\rho}\cdot\exp\pm{\rm i}\sqrt{2}\theta_{\sigma}$ &
 $ *$& $ 1$ & $ *$ & $0$ & $\pm 1$\\
 $4k_{\rm F}$-CDW &
 $\cos 2\sqrt{2}\phi_{\rho}$ &
 $ *$& $-1$ & $ *$ & $4k_{\rm F}$ & $\pm 1$\\
 \hline
 singlet &
 $\cos\sqrt{2}\phi_{\sigma}$ &
 $ 1$& $ 1$ & $ 1$ & $0$ & $\mp 1$\\
 triplet$_0$ &
 $\sin\sqrt{2}\phi_{\sigma}$ &
 $-1$& $-1$ & $-1$ & $0$ & $\mp 1$\\
 \ \ triplet$_{\pm1}$ &
 $\exp\pm{\rm i}\sqrt{2}\theta_{\sigma}$ &
 $ *$& $ 1$ & $ *$ & $0$ & $\mp 1$\\
 ``dimer'' &
 $\cos\sqrt{2}\phi_{\rho}$ &
 $ 1$& $ 1$ & $ *$ & $2k_{\rm F}$ & $\mp 1$\\
 ``N\'{e}el'' &
 $\sin\sqrt{2}\phi_{\rho}$ &
 $-1$& $-1$ & $ *$ & $2k_{\rm F}$ & $\mp 1$\\
 ``doublet'' &
 $\exp\pm{\rm i}\sqrt{2}\theta_{\rho}$ &
 $ *$& $ 1$ & $ 1$ & $0$ & $\mp 1$
\end{tabular}
\caption{Discrete symmetries of the excitation spectra (${\cal C}$:
 charge conjugation, ${\cal P}$: space inversion, ${\cal T}$: spin
 reversal, and $k$: wave number). BC$=1$ (BC$=-1$) stands for
 (anti)periodic boundary conditions. The upper (lower) sign of BC
 denotes $N/2=$ odd (even) cases, where $N$ is the number of electrons.
 The upper 12 states are ``physical'' ones, which appear under the same
 BC as those of the ground state. The lower six states are the
 ``artificial'' ones extracted by twisting BC with respect to the ground
 state.} \label{tbl:symmetries}
\end{table}

Next, we consider the symmetries of operators.  The operator of the
marginal field (\ref{eqn:marginal}) has the same quadratic form of the
Gaussian part of the Lagrangian density of Eq.~(\ref{eqn:eff_Ham}), so
that it has the same symmetry as the ground state (${\cal C}={\cal
P}={\cal T}=1$, $k=0$).  We can find the symmetries of the ${\cal
O}_{\nu 1}$ and the ${\cal O}_{\nu 2}$ operators by considering the
change of the phase fields. Since we restrict our attention to the
Hilbert space with fixed electron number and total $z$-spin, we do not
consider the change of the $\theta_{\nu}$ fields in the symmetry
operations.  At half-filling ($k_{\rm F}=\pi/2$), it follows from
Eq.~(\ref{eqn:fermion_op}) that the phase fields $\phi_{\nu}$ are
transformed under particle-hole (${\cal C}$:
$\psi_{r,s}\leftrightarrow\psi_{r,s}^{\dag}$), space-inversion (${\cal
P}$: R$\leftrightarrow$L, $x\rightarrow x+a$), spin-reversal
transformations (${\cal T}$: $\uparrow\leftrightarrow\downarrow$), and
shift operation (${\cal S}: x\rightarrow x+a$) as
\begin{mathletters}\label{eqn:CPTS}
\begin{eqnarray}
 {\cal C}:&&\phi_{\sigma}\rightarrow-\phi_{\sigma},\ \ \
            \phi_{\rho}\rightarrow-\phi_{\rho},\label{eqn:C}\\
 {\cal P}:&&\phi_{\sigma}\rightarrow-\phi_{\sigma},\ \ \
            \phi_{\rho}\rightarrow-\phi_{\rho},\label{eqn:P}\\
 {\cal T}:&&\phi_{\sigma}\rightarrow-\phi_{\sigma},\ \ \
            \phi_{\rho}\rightarrow\phi_{\rho},\label{eqn:T}\\
 {\cal S}:&&\phi_{\sigma}\rightarrow\phi_{\sigma},\ \ \ \ \
            \phi_{\rho}\rightarrow\sqrt{2}k_{\rm F}+\phi_{\rho}.\label{eqn:S}
\end{eqnarray}
\end{mathletters}
In this case, ${\cal C}{\cal P}=1$ is always satisfied, so that the
independent discrete symmetries are ${\cal P}$, ${\cal T}$, and ${\cal
S}$.  At quarter-filling ($k_{\rm F}=\pi/4$), the phase fields change as
\begin{equation}
 {\cal P}:\phi_{\sigma}\rightarrow-\phi_{\sigma},\ \ \
            \phi_{\rho}\rightarrow\pi/\sqrt{8}-\phi_{\rho}.
\end{equation}
Thus the discrete symmetries of the operator ${\cal O}_{\rho 1}$ for
$q=2$ are ${\cal P}=-1$.  The relations between the operators and their
symmetries are summarized in Table \ref{tbl:symmetries}.

In the present numerical calculation based on the Lancz\"{o}s algorithm,
the identification is performed by projecting the initial vector as
\begin{equation}
 |\Psi_{\rm init}\rangle=\frac{1}{2}(1\pm{\cal P})(1\pm{\cal T})|i\rangle,
\end{equation}
where the signs in front of the operators correspond to their
eigenvalues, and $|i\rangle$ is a configuration that satisfies ${\cal
P}, {\cal T}|i\rangle\neq|i\rangle$. Furthermore, $|i\rangle$ is
classified by the wave numbers $k=0,\pi$.

\section{phases}\label{sec:PHASES}
In this section, we discuss the character of each phase that appears in
the phase diagrams. In general, there are no long-range orders (LRO's)
in 1D systems due to strong quantum fluctuations, so that, in such
cases, the phases are characterized by the dominant correlation
functions.  The correlation functions (\ref{eqn:correlation_func})
including the logarithmic corrections are given by integrating the
renormalized scaling dimensions over the RG trajectory as
\begin{equation}
 R_i=\exp\left[-\int_0^{\ln(r/\alpha)}{\rm d}l\
	  2[x_{\rho i}(l)+x_{\sigma i}(l)]\right].
  \label{eqn:R_integral}
\end{equation}

First, we consider Eq.~(\ref{eqn:R_integral}) for the spin and charge
degrees of freedom independently.  For the spin part which has the SU(2)
symmetry, the singlet (${\cal O}_{\sigma 1}$) and the triplet (${\cal
O}_{\sigma 2,3}$) correlation functions with logarithmic corrections are
obtained explicitly in the gapless region ($g_{1\perp}^*=0,
K_{\sigma}^*=1$) as\cite{Giamarchi-S}
\begin{mathletters}
\begin{eqnarray}
 R_{\sigma 1}&=&
   \frac{\alpha}{r}\ln^{-3/2}(r/\alpha),\\
 R_{\sigma 2,3}&=&
  \frac{\alpha}{r}\ln^{1/2}(r/\alpha),
\end{eqnarray}
\end{mathletters}
where we have used Eqs.~(\ref{eqn:renormalized_coupling}),
(\ref{eqn:sclngdim_spin}), and (\ref{eqn:R_integral}).  Therefore, the
triplet correlation is more logarithmically dominant than the singlet
one.  On the other hand, when the spin gap opens
($g_{1\perp}^*=-\infty$), the singlet excitation degenerates with the
ground state in the thermodynamic limit, so that the singlet correlation
becomes constant, while the triplet one decays exponentially.  In this
way, the triplet correlation is suppressed in the spin-gap region.

For the charge part, at half-filling, explicit forms of the correlation
functions including logarithmic corrections are not obtained except for
the BKT or the Gaussian lines.  On the BKT line, the exponent is
renormalized as $K_{\rho}^{*}=1$, so that $K_{\rho}^{*}\geq 1$ is always
satisfied in the gapless region ($g_{3\perp}^*=0$), and then the
correlation for the ``doublet'' (${\cal O}_{\rho 3}$) is dominant.  In
the charge-gap region with $g_{3\perp}^*=\infty$, the ``N\'{e}el''
(${\cal O}_{\rho 2}$) state degenerates with the ground state, and the
``dimer'' (${\cal O}_{\rho 1}$) and the ``doublet'' correlations decay
exponentially.  On the other hand, for $g_{3\perp}^*=-\infty$, the
``dimer'' state degenerates with the ground state, and the ``N\'{e}el''
and the ``doublet'' correlations decay exponentially.

Next, we discuss the physical states that consist of the charge and the
spin parts.  In the metallic region ($g_{3\perp}^{*}=0, K_{\rho}^{*}\geq
1$), the triplet superconducting (TS) correlation is dominant when the
spin part is gapless.  The operators for the TS phase consist of the
``doublet'' and the triplet ones,
\begin{mathletters}
\begin{eqnarray}
 {\cal O}_{{\rm TS}_{0}}&=&
  \sum_{s} c^{\dag}_{js}c^{\dag}_{j+1,-s},\nonumber\\
 &\propto&
 \exp[+{\rm i}\sqrt{2}\theta_{\rho}(x)]\sin[\sqrt{2}\phi_{\sigma}(x)],\\
 {\cal O}_{{\rm TS}_{1}}&=&
  c^{\dag}_{j\uparrow}c^{\dag}_{j+1,\uparrow},\nonumber\\
 &\propto&
 \exp[+{\rm i}\sqrt{2}\theta_{\rho}(x)]
 \exp[+{\rm i}\sqrt{2}\theta_{\sigma}(x)].
 \label{eqn:TS1}
\end{eqnarray}
\end{mathletters}
On the other hand, the singlet superconducting (SS) correlation whose
operator is given by the ``doublet'' and the singlet ones
\begin{eqnarray}
 {\cal O}_{\rm SS}&=&c^{\dag}_{j\uparrow}c^{\dag}_{j\downarrow},\nonumber\\
 &\propto&
  \exp[+{\rm i}\sqrt{2}\theta_{\rho}(x)]\cos[\sqrt{2}\phi_{\sigma}(x)],
\end{eqnarray}
is dominant when the spin part has a gap.

In the insulating region, which corresponds to the fixed point
$g_{3\perp}^{*}=+\infty$, the bond-spin-density-wave (BSDW)
phase\cite{Japaridze} characterized by
\begin{mathletters}\label{eqn:BSDW}
\begin{eqnarray}
 {\cal O}_{{\rm BSDW}\alpha}&=&(-1)^j\sum_{s,s'}
(c^{\dag}_{js}\tau^{\alpha}_{ss'}c_{j+1,s'}+
 c^{\dag}_{j+1,s}\tau^{\alpha}_{ss'}c_{js'}),\nonumber\\
 {\cal O}_{{\rm BSDW}z}
 &\propto&
 \sin[\sqrt{2}\phi_{\rho}(x)]\sin[\sqrt{2}\phi_{\sigma}(x)],
 \label{eqn:BSDWz}\\
 {\cal O}_{{\rm BSDW}\pm}
 &\propto&
 \sin[\sqrt{2}\phi_{\rho}(x)]\exp[\pm{\rm i}\sqrt{2}\theta_{\sigma}(x)],
 \label{eqn:BSDW+}
\end{eqnarray}
\end{mathletters}
appears when the spin sector is gapless. On the other hand, the
charge-density-wave (CDW) phase
\begin{eqnarray}
 {\cal O}_{\rm CDW}&=&(-1)^j\sum_s c^{\dag}_{js}c_{js},\nonumber\\
 &\propto&
  \sin[\sqrt{2}\phi_{\rho}(x)]\cos[\sqrt{2}\phi_{\sigma}(x)],
 \label{eqn:CDW}
\end{eqnarray}
appears when the spin gap opens. In the CDW phase, both charge and spin
gaps open, so that a LRO exists.

\begin{figure}[h]
\epsfxsize=2.8in \leavevmode \epsfbox{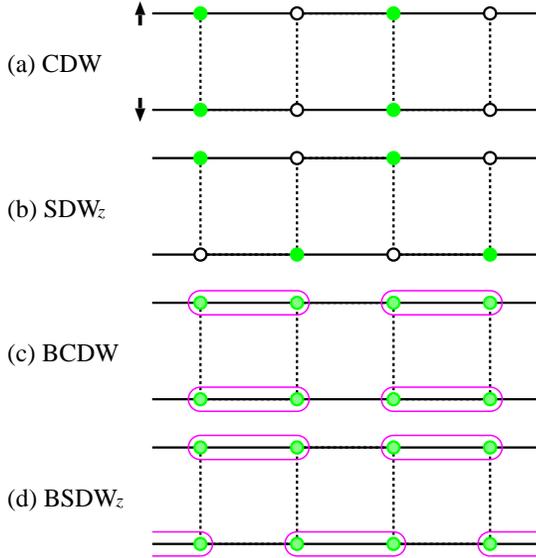}
\caption{Schematic illustration for the four charge-gapped states in up-
 and down-spin subsystems.  The enclosed two sites in (c) and (d) stand
 for electron-hole dimers.  The electrons polarize on sites (${\cal
 C}={\cal P}=-1$) for CDW and SDW states, while they polarize on bonds
 (${\cal C}={\cal P}=+1$) for BCDW and BSDW states. The two subsystems
 are synchronized (${\cal T}=+1$) for CDW and BCDW states, while they
 are displaced by one site (${\cal T}=-1$) for SDW and BSDW states.}
 \label{fig:ladders}
\end{figure}
\begin{table}[H]
\begin{tabular}{c|c|c|c}
          & $g_{3\perp}^*=0$
          & $g_{3\perp}^*=+\infty$
          & $g_{3\perp}^*=-\infty$\\\hline
 $g_{1\perp}^*=0$
       & TS & BSDW & SDW\\\hline
 $g_{1\perp}^*=-\infty$
 & SS & CDW (LRO)& BCDW (LRO)
\end{tabular}
\caption{Correspondence between six possible phases at half-filling and
 fixed points in the RG analysis.  For $g_{1\perp}^*=-\infty$
 ($g_{1\perp}^*= 0$), the spin sector is gapped (gapless).  For
 $g_{3\perp}^*=\pm\infty$ ($g_{3\perp}^*=0$), the charge sector is
 gapped (gapless).}  \label{tbl:charge-gapped-phases}
\end{table}

In the insulating region for the opposite fixed point
$g_{3\perp}^{*}=-\infty$, the spin-density-wave (SDW) correlation
characterized by
\begin{mathletters}\label{eqn:SDW}
\begin{eqnarray}
 {\cal O}_{{\rm SDW}\alpha}&=&
  (-1)^j\sum_{s,s'} c^{\dag}_{js}\tau^{\alpha}_{ss'}c_{js'},\nonumber\\
 {\cal O}_{{\rm SDW}z}
 &\propto&
\cos[\sqrt{2}\phi_{\rho}(x)]\sin[\sqrt{2}\phi_{\sigma}(x)],
 \label{eqn:SDWz}\\
 {\cal O}_{{\rm SDW}\pm}
 &\propto&
 \cos[\sqrt{2}\phi_{\rho}(x)]\exp[\pm{\rm i}\sqrt{2}\theta_{\sigma}(x)],
 \label{eqn:SDW+}
\end{eqnarray}
\end{mathletters}
is dominant when the spin part is gapless. The bond-charge-density-wave
(BCDW) phase characterized by
\begin{eqnarray}
 {\cal O}_{\rm BCDW}&=&
  (-1)^j \sum_s(c^{\dag}_{js}c_{j+1,s}+c^{\dag}_{j+1,s}c_{js}),\nonumber\\
 &\propto&
 \cos[\sqrt{2}\phi_{\rho}(x)]\cos[\sqrt{2}\phi_{\sigma}(x)].
 \label{eqn:BCDW}
\end{eqnarray}
appears when the spin gap opens. In the BCDW phase, both charge and spin
gaps open, so that a LRO exists.  The correspondence between the above
six phases at half-filling and the fixed points are summarized in Table
\ref{tbl:charge-gapped-phases}.

At quarter-filling ($q=2$), the $4k_{\rm F}$-charge-density wave
($4k_{\rm F}$-CDW) appears when the Umklapp scattering is relevant. The
operator is given by the ${\cal O}_{\rho 1}$ field with $q=2$,
\begin{eqnarray}
   {\cal O}_{\rm 4CDW}&=&
 \sum_{r}\psi_{r\uparrow}^{\dag}(x)\psi_{r\downarrow}^{\dag}(x)
  \psi_{-r\downarrow}(x)\psi_{-r\uparrow}(x),\nonumber\\
 &\propto&\cos[\sqrt{8}\phi_{\rho}(x)].
\end{eqnarray}

In the rest of this section, we further clarify the differences among
the four charge-gapped states (CDW, SDW, BCDW, and BSDW) discussed
above.  For this purpose, we change the basis of the bosonized operators
from the charge and spin picture to the spin up and down one by
introducing the following new phase fields:
\begin{equation}
 \phi_{s}=\phi_{\rho}\pm\phi_{\sigma},\label{eqn:new_fields}
\end{equation}
where $s=\uparrow,\downarrow$ refer to the upper and lower signs,
respectively. Then, the system is interpreted as coupled spinless
fermion systems ($S=1/2$ spin chains).  In this case, the (B)CDW and
$z$-components of the (B)SDW operators are given by
\begin{eqnarray}
 {\cal O}_{\rm CDW}, {\cal O}_{{\rm SDW}_{z}}
  &\propto&
  \sin(\sqrt{2}\phi_{\downarrow})\pm\sin(\sqrt{2}\phi_{\uparrow}),
  \label{eqn:CDW&SDW}\\
 {\cal O}_{\rm BCDW}, {\cal O}_{{\rm BSDW}_{z}}
  &\propto&
  \cos(\sqrt{2}\phi_{\uparrow})\pm\cos(\sqrt{2}\phi_{\downarrow}),
  \label{eqn:BCDW&BSDW}
\end{eqnarray}
where the CDW and the BCDW (the SDW$_{z}$ and the BSDW$_{z}$) operators
refer to the upper (lower) signs in the right-hand sides, and
$\sin(\sqrt{2}\phi_s)$ and $\cos(\sqrt{2}\phi_s)$ fields denote N\'{e}el
and dimer states of the $S=1/2$ spin chains, respectively.  In addition,
it follows from Eqs.~(\ref{eqn:S}) and (\ref{eqn:new_fields}) that the
shift operation by one site gives
\begin{equation}
 \phi_s\rightarrow\phi_s+\pi/\sqrt{2},
\end{equation}
so that both CDW and SDW$_{z}$ states are described by two N\'{e}el
ordered spin chains, where the two sectors ($s=\uparrow,\downarrow$) are
synchronized in the former, while they are displaced by one site in the
latter.  On the other hand, the BCDW and the BSDW$_{z}$ states are given
by synchronized and displaced dimer-ordered spin chains,
respectively (see Fig.~\ref{fig:ladders}).  Therefore, in
the BCDW phase, the charge is polarized on the bonds alternatively, and
the spins are dimerized.  In the BSDW state, the charge is polarized on
the each bond, and the spins are located on the bonds and remain gapless
like the SDW state.\cite{Japaridze}

The discrete symmetries discussed in Sec.~\ref{sec:SYMMETRY}
characterize the differences among these physical states. It follows
from Eq.~(\ref{eqn:CPTS}), that the spin reversal symmetry corresponds
to whether $s=\uparrow,\downarrow$ sectors are synchronized (${\cal
T}=+1$) or displaced by one site (${\cal T}=-1$). Similarly, the parity
and the charge conjugation distinguish whether the electrons are
polarized on the sites (${\cal C}={\cal P}=-1$) or on the bonds (${\cal
C}={\cal P}=+1$).  This interpretation for the parity is consistent with
the fact that the $4k_{\rm F}$-CDW with a site LRO has the odd parity
(${\cal P}=-1$) at quarter-filling.

\section{HALF-FILLING}\label{sec:HALF}
Using the method explained in Sec.~\ref{sec:theory}, we analyze the
phase diagram of the EHM at half-filling. Since there are many
instabilities, we consider the phase diagram separately in the charge
and the spin parts by assuming the charge-spin separation. However, for
the CDW-SDW transition, there is a possibility that the charge and the
spin degrees of freedom are coupled. So we discuss two cases where the
two degrees of freedom are separated and coupled. And then we identify
the valid scenario from the comparison with the result of the
strong-coupling perturbation theory. For the phase separation, which is
considered to be a first-order transition, we need an approach different
from the one applied to the charge- and the spin-gap phases.  We discuss
the way to determine the phase-separation boundary and check the
validity of the result by the strong-coupling theory.  Using the above
strategy, we also analyze the EHM with the correlated hopping term
[Eq.~(\ref{eqn:X-term})].  The results are summarized in
Fig.~\ref{fig:PD1}.

\subsection{Spin sector}
\begin{figure}[h]
\noindent
\epsfxsize=3.5in \leavevmode \epsfbox{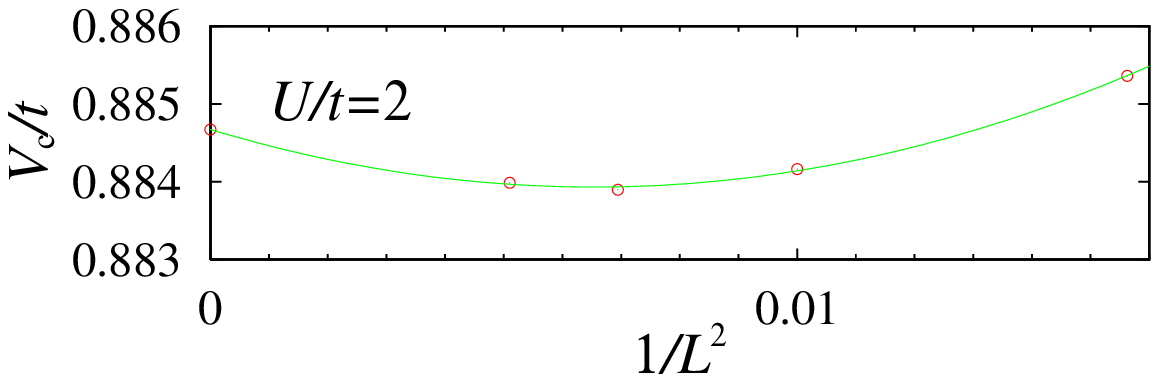}
\caption{Size dependence of the critical point for the spin-gap
 transition at $U/t=2$.  The system sizes are $L=8,10,12,14$.
 From Ref.~\protect{\ref{Nakamura_99}}.}
 \label{fig:size_sg}
\noindent 
\epsfxsize=3.5in \leavevmode \epsfbox{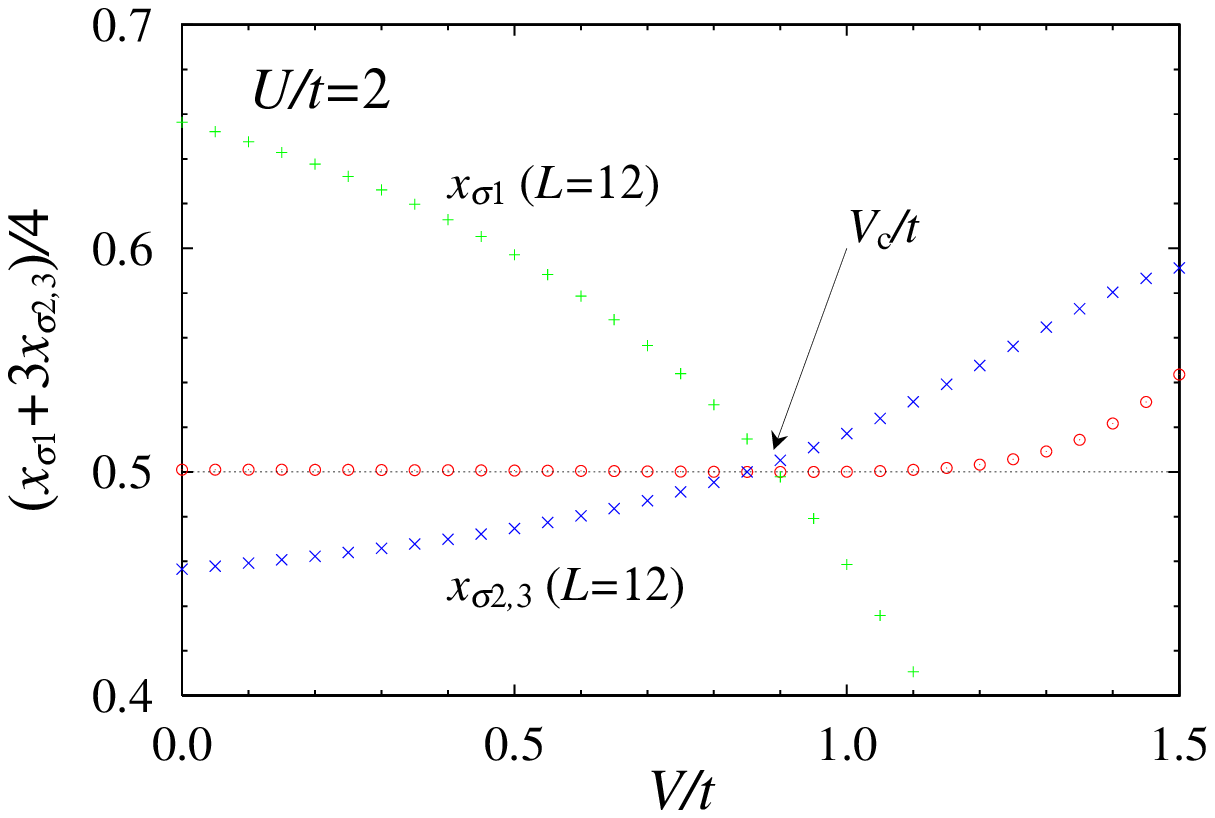}
\caption{Extrapolated scaling dimension given by
 Eq.~(\protect{\ref{eqn:check_sg}}) near the spin-gap critical point at
 $U/t=2$. The TL liquid theory predicts the numerical values are $1/2$
 in the gapless region ($V<V_{\rm c}$). This result shows the existence
 of the spin-gap transition.  The scaling dimensions for the singlet
 ($x_{\sigma 1}$) and the triplet ($x_{\sigma 2,3}$) excitation spectra
 in the $L=12$ system are also shown.}  \label{fig:check_sg}
\end{figure}

First, we determine the spin-gap phase boundary following the method
explained in Sec.~\ref{sec:spin-gap}.  By observing the singlet-triplet
level crossing, the phase boundary is found to be near the $U=2V$ line.
Since the critical point is almost independent of the system size (see
Fig.~\ref{fig:size_sg}), we can determine the phase boundary without any
extrapolations.  In order to check the consistency of our argument, we
calculate scaling dimensions of the singlet and the triplet excitations
from Eq.~(\ref{eqn:energy}), and confirm the ratios of the logarithmic
corrections.  Here the spin-wave velocity is given by the excitation
spectra for $N_{\sigma}(\bar{N}_{\sigma})=1$ or
$|n_{\sigma}|=|m_{\sigma}|=1$ as
\begin{equation}
 v_{\sigma}=\lim_{L\rightarrow\infty}
  \frac{E(L,N,S=1,k=2\pi/L)-E_0(L,N)}{2\pi/L},
\end{equation}
where the extrapolation is performed by the function $v_{\sigma}(L) =
v_{\sigma}(\infty) + A/L^2 + B/L^4$, which is explained by $x_{\nu}=4$
irrelevant fields.\cite{Cardy86,Reinicke}  Physically, this correction
is related to the deviation from the linearized dispersion relation
assumed in the TL model.  Thus, the ratio of the logarithmic corrections
can be checked as $3:-1$ for the singlet and the triplet states by using
the following relation near the critical point,
\begin{equation}
  \frac{x_{\sigma 1}+3x_{\sigma 2,3}}{4}=\frac{1}{2}. 
   \label{eqn:check_sg}
\end{equation}
Here we use the numerical data of $L=8,10,12$ systems to check the
scaling dimensions, and extrapolate them by the form $A+BL^{-2}+CL^{-4}$
as in the same way of the spin velocity.  As shown in
Fig.~\ref{fig:check_sg}, the extrapolated data become $1/2$ in the
gapless region.  Thus, the universality of the transition is identified
as the level-1 SU(2) WZNW model.

\subsection{Charge sector}\label{ssec:Mott}
\begin{figure}[b]
\noindent
\epsfxsize=3.4in \leavevmode \epsfbox{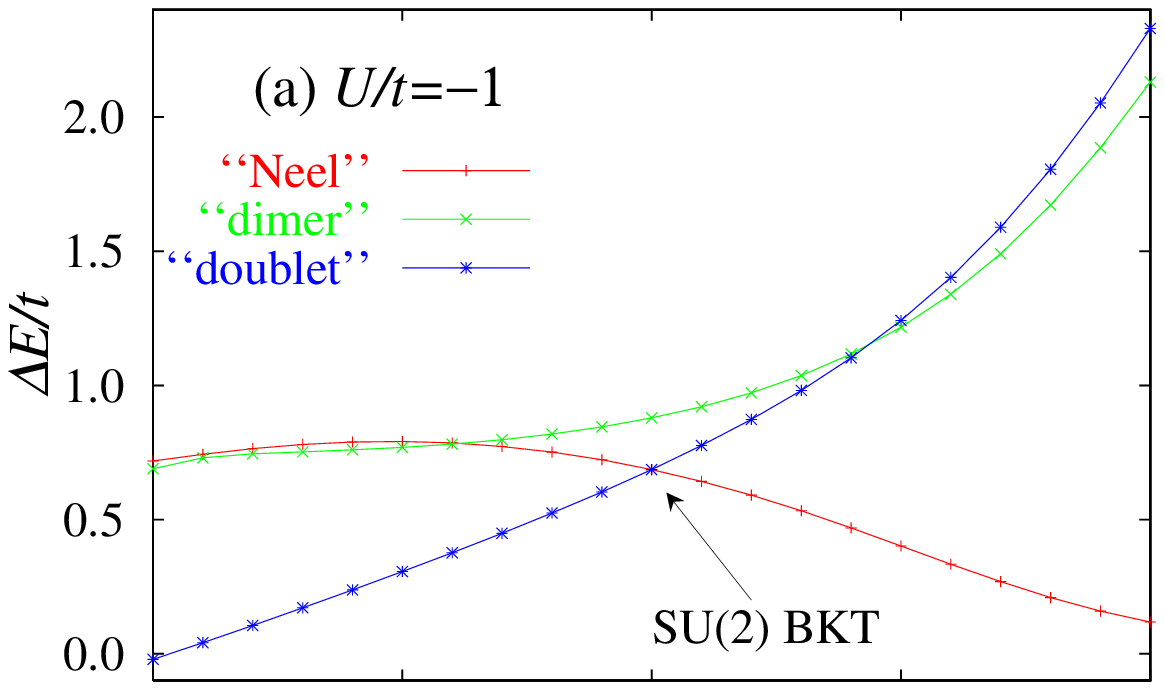}\\
\epsfxsize=3.4in \leavevmode \epsfbox{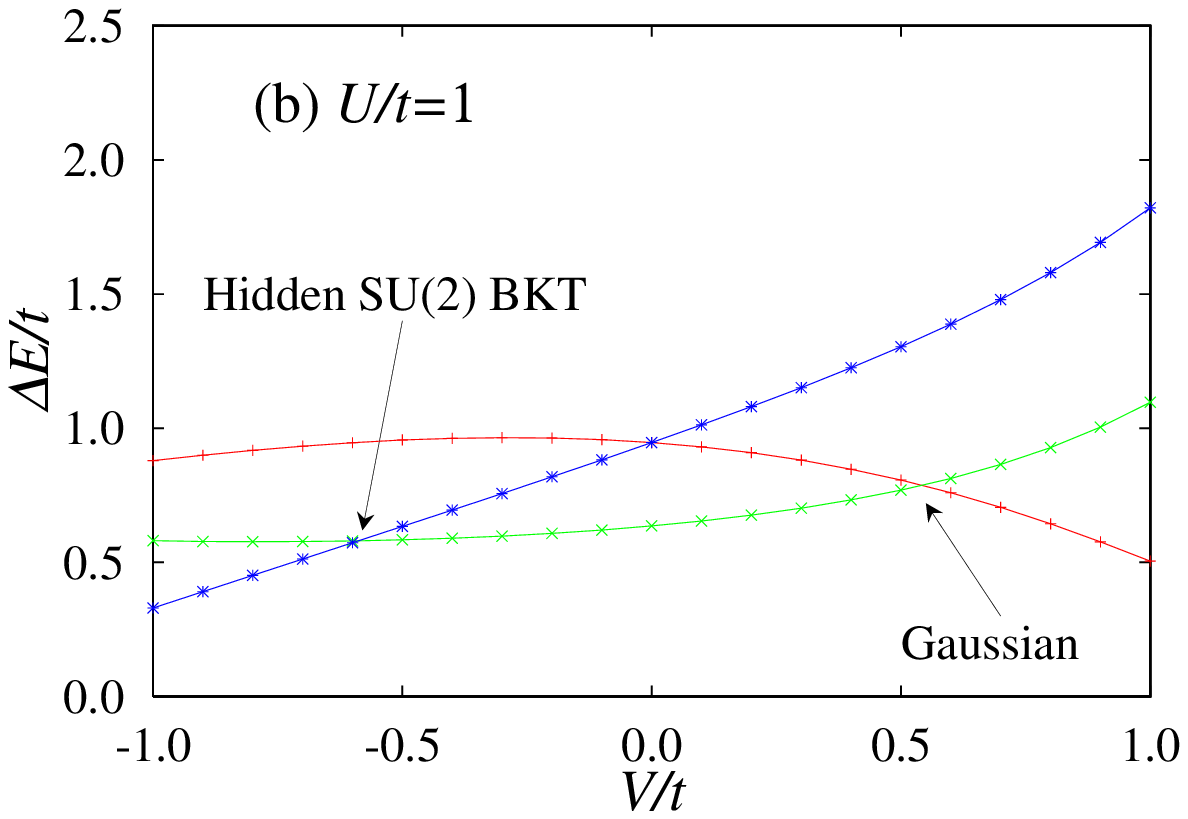}
\caption{``Dimer''($x_{\rho 1}$), ``N\'{e}el''($x_{\rho 2}$), and
``doublet''($x_{\rho 3}$) excitation spectra in the charge sector vs
$V/t$ in $L=8$ system at (a) $U/t=-1$ and (b) $U/t=1$. In the $U/t<0$
region, a BKT-type transition takes place at $V/t=0$ reflecting the
SU(2) symmetry of the $\eta$-pairing. In the $U/t>0$ region, two level
crossings occur due to the hidden SU(2) symmetric BKT and the Gaussian
transitions. These three level crossings give the Y-shaped structure in
the phase diagram.}  \label{fig:LEVEL_CROSS}
\end{figure}

Due to the SU(2) symmetry of the $\eta$-pairing (\ref{eqn:eta-pairing}),
a BKT transition takes place on the $V=0$ line for $U<0$ region. This
phase boundary is fixed on this line for any strength of $U<0$. In
Fig.~\ref{fig:LEVEL_CROSS}(a), the degeneracy of the ``N\'{e}el''
($x_{\rho 2}$) and the ``doublet'' ($x_{\rho 3}$) excitation spectra on
$V=0$ corresponds to this SU(2) symmetry.  On the other hand, for the
$U>0$ region, there appear two relevant level crossings as shown in
Fig.~\ref{fig:LEVEL_CROSS}(b). The one corresponds to the BKT transition
due to the hidden SU(2) symmetry, and the other corresponds to the
Gaussian transition, as was explained in Sec.~\ref{sec:theory}.  The
rest of the three level crossings in Fig.~\ref{fig:LEVEL_CROSS} do not
correspond to any phase transitions, because they correspond to the
lines $\pm y_{\rho\phi}(0)=y_{\rho 0}(0)<0$ and the line
$y_{\rho\phi}(0)=0$ with $y_{\rho 0}(0)>0$ (Gaussian fixed line), in the
RG flow diagram of Fig.~\ref{fig:RGflow}(b).

\begin{figure}
\noindent
\epsfxsize=3.4in \leavevmode \epsfbox{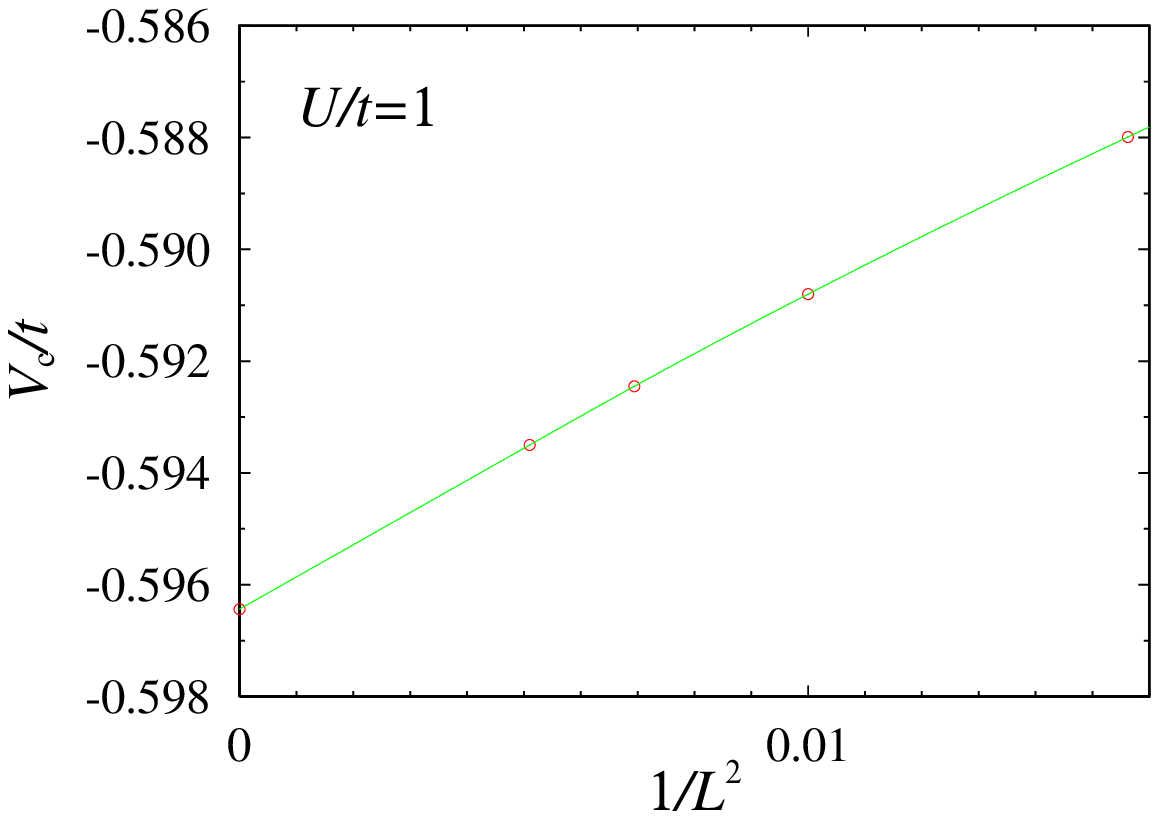}
\caption{Size dependence of the critical point of the BKT transition due
 to the hidden SU(2) symmetry at $U/t=1$. The system sizes are
 $L=8,10,12,14$.}
 \label{fig:SIZE_BKT}
\epsfxsize=3.3in \leavevmode \epsfbox{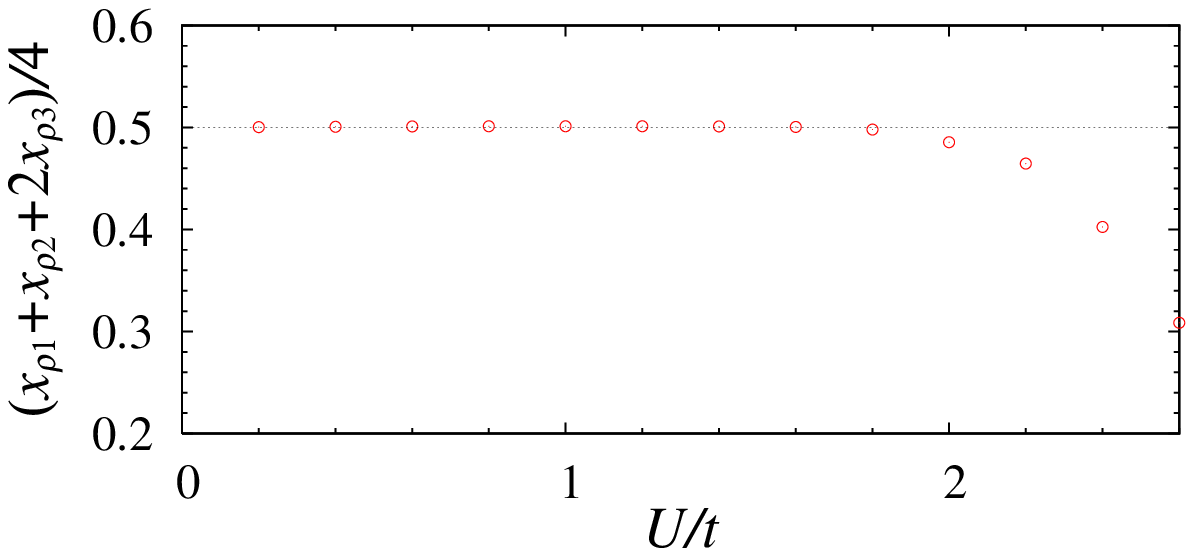}
\caption{Extrapolated scaling dimension given by
Eq.~(\protect{\ref{eqn:check_cg}}) on the BKT critical line. The TL
liquid theory predicts the numerical values are $1/2$. This result shows
the existence of the BKT-type transition.}
 \label{fig:check_cg}
\end{figure}

The hidden SU(2) BKT transition obtained by the ``dimer''-``doublet''
level crossing ($x_{\rho 1}=x_{\rho 3}$) appears near the $U=-2V$ line
as was predicted by the g-ology. The size dependence of this transition
at $U/t=1$ is shown in Fig.~\ref{fig:SIZE_BKT}. In order to check the
consistency of our argument, we calculate the scaling dimensions of
$x_{\rho i}$ using Eq.~(\ref{eqn:energy}). Here, we calculate the charge
velocity using the excitation spectra for $N_{\rho}(\bar{N}_{\rho})=1$
as
\begin{equation}
 v_{\rho}=\lim_{L\rightarrow\infty}
  \frac{E(L,N,S=0,k=2\pi/L)-E_0(L,N)}{2\pi/L}.
\end{equation}
Using Eqs.~(\ref{eqn:sclngdim_charge_BKT}), we check the following
relation on the critical line,
\begin{equation}
 \frac{x_{\rho 1}+x_{\rho 2}+2x_{\rho 3}}{4}=\frac{1}{2}.
  \label{eqn:check_cg}
\end{equation}
As shown in Fig.~\ref{fig:check_cg}, the extrapolated data become $1/2$.
Here, the extrapolation is performed as in the same way of the spin-gap
transition.  Thus, the universality class of this transition is
identified as a BKT type. The deviation from the expected value $1/2$ in
Fig.~\ref{fig:check_cg} stands for the effect of the phase separation
where the TL liquid theory breaks down.

\begin{figure}
\noindent \epsfxsize=3.5in \leavevmode \epsfbox{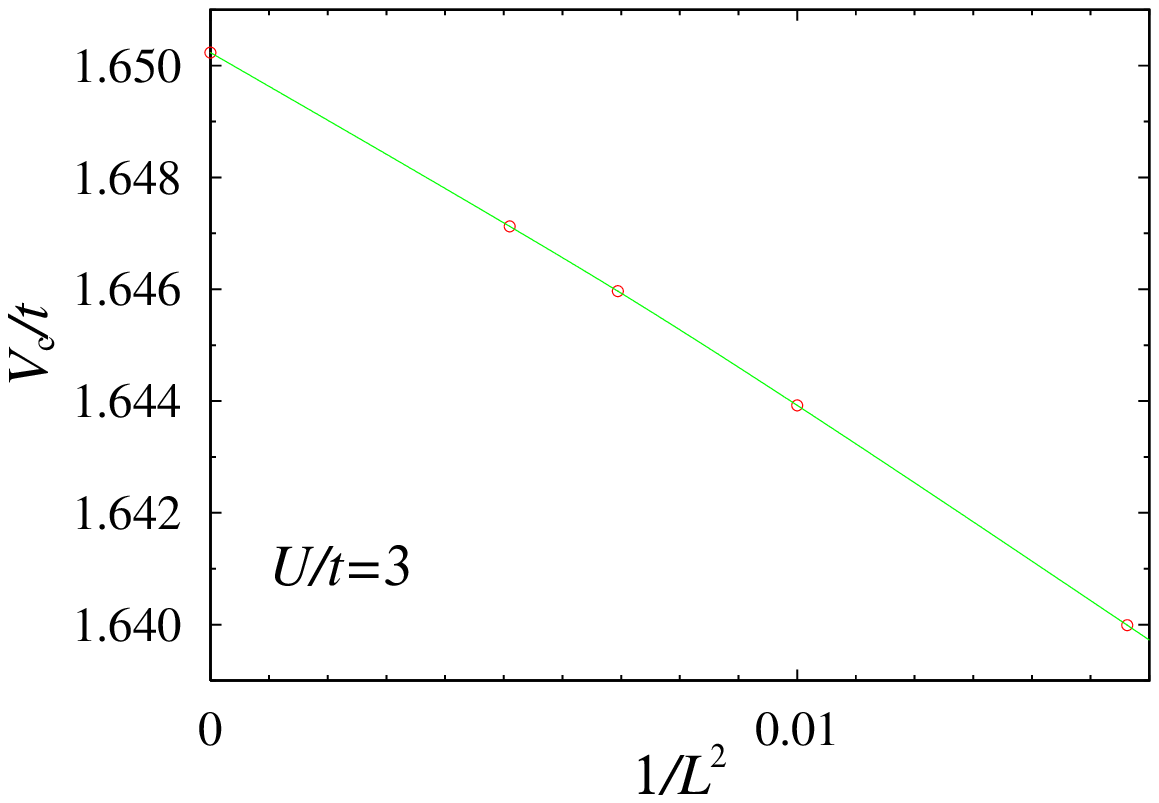}
\caption{Size dependence of the critical point for the Gaussian
transition at $U/t=3$.  The system sizes are $L=8,10,12,14$.  The data
agrees with the result of Cannon {\it et
al.}\protect{\cite{Cannon-S-F}} that $V_{\rm
c}/t=1.65^{+0.10}_{-0.05}$.  From Ref.~\protect{\ref{Nakamura_99}}.}
\label{fig:size_ga}
\noindent
\epsfxsize=3.5in \leavevmode \epsfbox{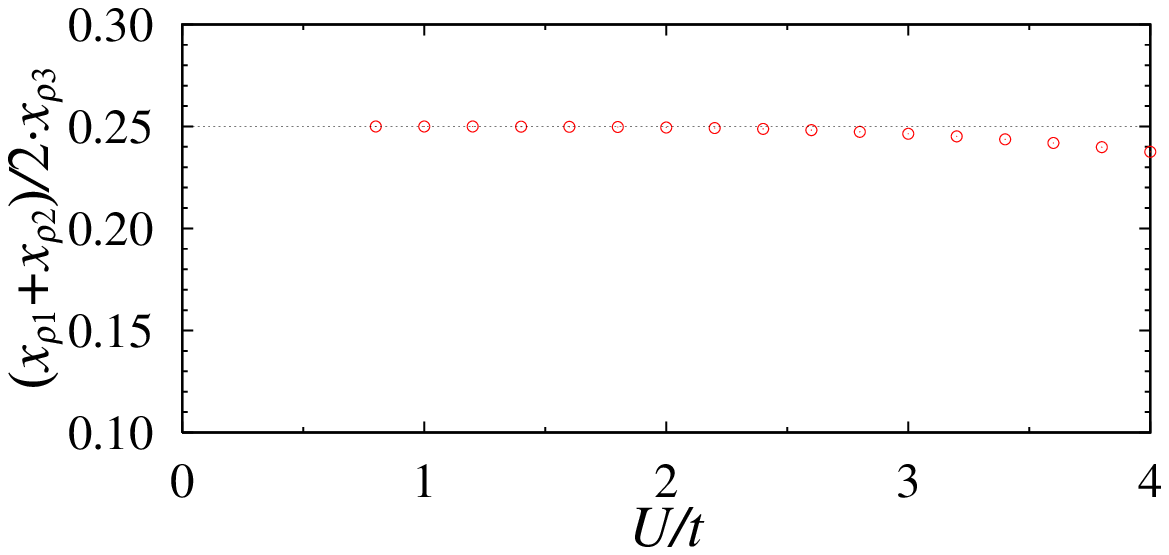}
\caption{Extrapolated scaling dimension given by
 Eq.~(\protect{\ref{eqn:check_ga}}) on the Gaussian critical line.  The
 TL liquid theory predicts the numerical values are $1/4$.  This result
 shows the existence of the Gaussian transition.}
 \label{fig:check_ga}
\end{figure}

The Gaussian transition takes place along the $U=2V$ line as was
predicted by the g-ology. The size dependence of this transition at
$U/t=3$ is shown in Fig.~\ref{fig:size_ga}.  It follows from
Eq.~(\ref{eqn:sclngdim_charge_gaussian}), the following relation should
be satisfied just on the Gaussian transition line,
\begin{equation}
 \frac{x_{\rho 1}+x_{\rho 2}}{2} x_{\rho 3}=\frac{1}{4}.
 \label{eqn:check_ga}
\end{equation}
The result is shown in Fig.~\ref{fig:check_ga}.  The extrapolated data
become $1/4$ from the weak- to the intermediate-coupling region.  Thus,
the transition is identified as a Gaussian type except for the
strong-coupling region.

\subsection{Transition between CDW and SDW phases}

We have determined the spin-gap and Gaussian transition lines near the
$U=2V$ line assuming the charge-spin separation, however, if the
$g_{3\parallel}$ term in Eq.~(\ref{eqn:eff_Ham}) is relevant, the
charge-spin coupling may take place.  Therefore, we should consider the
possibility that the charge and the spin degrees of freedom are not
separated, and that a direct transition between the CDW and the SDW
phases takes place. To examine this possibility, we also observe the
level crossing of excitation spectra of the CDW and the SDW operators
(see Table \ref{tbl:symmetries}), which consist of both charge and spin
components. These spectra can be obtained under conditions given in
Table \ref{tbl:symmetries}.

\begin{figure}[h]
\noindent
\epsfxsize=3.4in \leavevmode \epsfbox{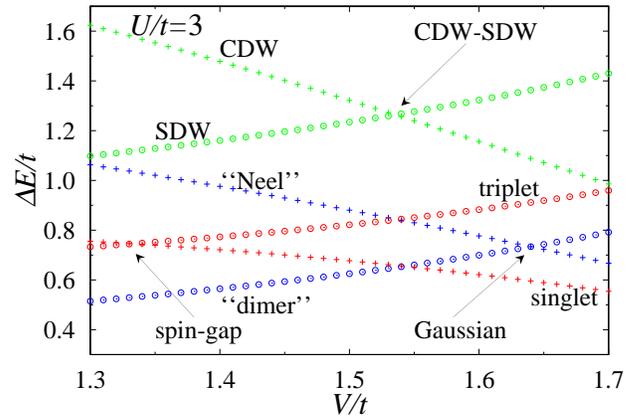}
\caption{Six excitation spectra vs $V/t$ in $L=8$ system at $U/t=3$
 near the $U=2V$ line.}
 \label{fig:LEVEL_CROSS2}
\end{figure}

\begin{figure}
\noindent
\epsfxsize=3.3in \leavevmode \epsfbox{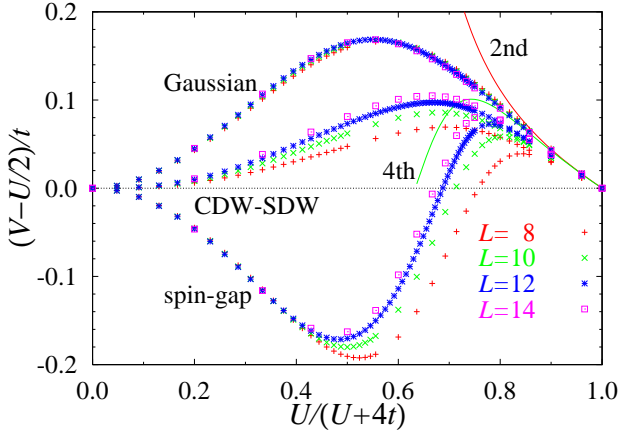}
\caption{Three possible transitions (Gaussian, spin-gap, and direct
 CDW-SDW transitions) along the $U=2V$ line of the EHM calculated in
 $L=8,10,12,14$ systems. The result of the strong-coupling expansion
 [Eq.~(\protect{\ref{eqn:CDW-SDW}}) with $\xi=0$] agrees with the
 Gaussian transition. This means that the actual transitions are the
 Gaussian and the spin-gap transitions, and a BCDW state exists between
 them.  From Ref.~\protect{\ref{Nakamura_99}}.}  \label{fig:3lines}
\end{figure}

The level-crossing points for the three assumed transition lines are
close to the $U=2V$ line, but slightly deviate (see
Fig.~\ref{fig:LEVEL_CROSS2}). The deviations from $U=0$ to $U=\infty$
are shown in Fig.~\ref{fig:3lines}. These three lines coincide in the
weak- and the strong-coupling limits.  For the Gaussian line, the
finite-size effect is small for all regions. For the spin-gap phase
boundary, the finite-size effect is small in the weak-coupling region,
but it becomes large in the strong-coupling region. The direct CDW-SDW
level-crossing point has large size effect for all regions.

In order to identify the actual transition lines from these three lines,
we use the strong-coupling perturbation theory following
Hirsch\cite{Hirsch} and van Dongen.\cite{Dongen} The energy of the SDW
state in the strong-coupling region of the EHM are analytically obtained
up to the fourth order.  If we include the correlated hopping term
(\ref{eqn:X-term}) in the EHM, the energies of the CDW and the SDW
states are given by
\begin{eqnarray}
 \frac{E_{\rm CDW}}{L}&=&
\frac{U}{2}-\frac{2(1-\xi)^2t^2}{(3v-1)U}\label{eqn:E_CDW}\\
&&\hspace{-1cm}
 +\frac{(1-\xi)^2\left[(36v^2-5v-1)(1-\xi)^2-8(3v-1)v\right]t^4}
{v(3v-1)^3(4v-1)U^3},\nonumber\\
 \frac{E_{\rm SDW}}{L}&=&
 V-\frac{4(1-\xi)^2t^2\ln 2}{(1-v)U}\label{eqn:E_SDW}\\
&&
 +9\zeta(3)\frac{(1-\xi)^2\left[2(1-\xi)^2-1+v\right]t^4}
  {(1-v)^3U^3},
\nonumber
\end{eqnarray}
where $v\equiv V/U$ and $\xi\equiv X/t$. In Eq.~(\ref{eqn:E_SDW}), we
have used the Bethe-ansatz result of the $S=1/2$ Heisenberg spin
chain:\cite{Hulthen,Takahashi}
\begin{mathletters}
\begin{eqnarray}
 \langle{\bm S}_i\cdot{\bm S}_{i+1}\rangle-{\textstyle\frac14}
  &=&-\ln 2,\\
 \langle{\bm S}_i\cdot{\bm S}_{i+2}\rangle-{\textstyle\frac14}
  &=&-4\ln 2+{\textstyle\frac94}\zeta(3).
\end{eqnarray}
\end{mathletters}
The phase boundary between the CDW and the SDW phases is given by the
equation\cite{Dongen_comment}
\begin{equation}
 E_{\rm CDW}=E_{\rm SDW}.\label{eqn:CDW-SDW}
\end{equation}
The strong-coupling theory shows a good agreement with the Gaussian
transition in the charge part, among the three transition lines that we
have considered.  We should also note that the present Gaussian critical
point agrees with Cannon {\it et al.}'s result obtained by the direct
evaluation of the CDW order parameter:\cite{Cannon-S-F} $V_{\rm
c}/t=1.65^{+0.10}_{-0.05}$ for $U/t=3$, and $V_{\rm c}/t=2.92\pm 0.04$
for $U/t=5.5$ (see Figs.~\ref{fig:size_ga} and \ref{fig:3lines}).

From the above results, we conclude that the actual transition near the
$U=2V$ line is not a direct CDW-SDW transition, but two independent
Gaussian and spin-gap transitions, at least from the weak- to the
intermediate-coupling region. In the strong-coupling region, these two
boundaries approach and appear to be coupled at finite $U$ and $V$.
Unfortunately, in the present analysis, we cannot determine this
tricritical point, but it is considered to be identical to the crossover
point between the second- and the first-order transitions.  This
phenomenon is considered to be an effect of the charge-spin coupling
term [the $g_{3\parallel}$ term in Eq.~(\ref{eqn:eff_Ham})] as was
discussed by Voit in Ref.~\ref{Voit92}.  In this way, our analysis
suggests that the crossover along the $U=2V$ line is closely related to
the validity of the charge-spin separation.  Our result also
demonstrates that there is a finite region of a charge- and spin-gapped
state between the Gaussian and the spin-gap transition lines.  It
follows from the discussion in Sec.~\ref{sec:PHASES} that the third
phase is identified as a BCDW state.

\begin{figure}
\noindent
\epsfxsize=1.65in \leavevmode \epsfbox{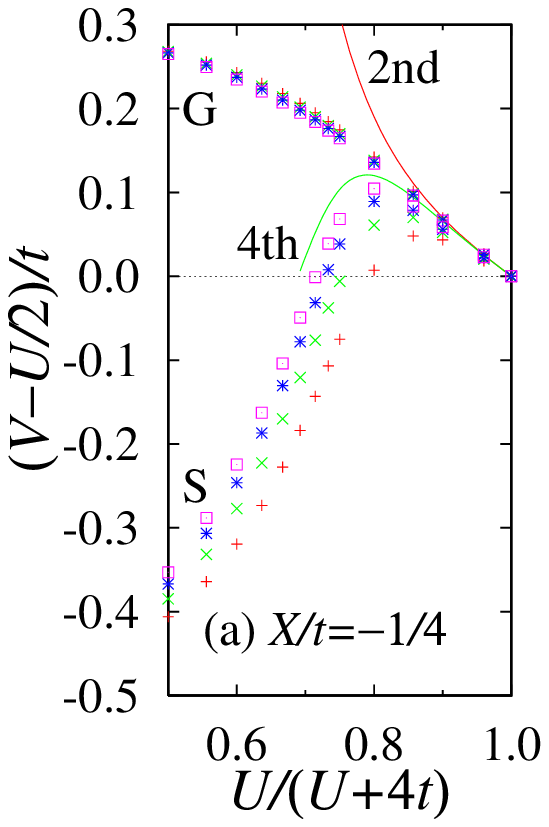}
\epsfxsize=1.65in \leavevmode \epsfbox{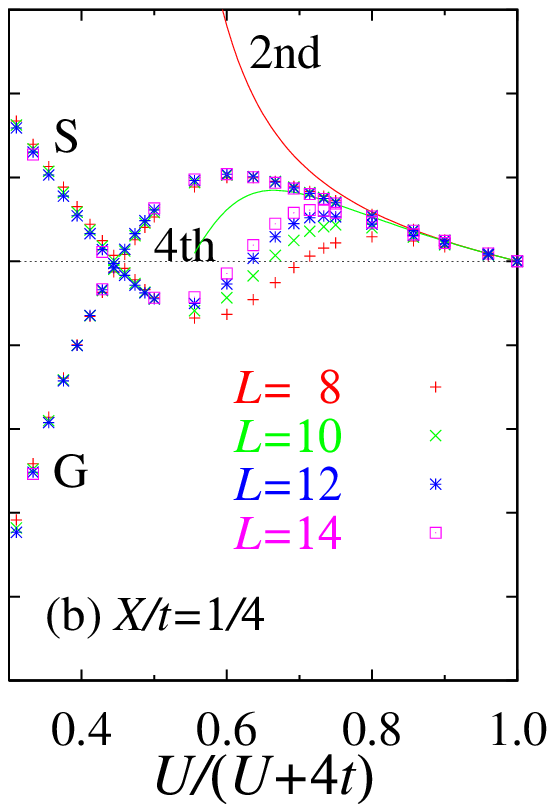}
\caption{The Gaussian (G) and the spin-gap (S) transitions in the
 strong-coupling region for (a) $X/t=-1/4$ and (b) $X/t=1/4$ . The
 strong-coupling theory [Eq.~(\protect{\ref{eqn:CDW-SDW}})] agrees with
 the Gaussian lines. Note that a BCDW phase appears not only in (a) but
 also in (b).}  \label{fig:diffs}
\end{figure}

In order to clarify the above interpretation for the CDW-SDW transition,
we analyze the EHM including the correlated hopping term
[Eq.~(\ref{eqn:X-term})]. This term is known to enlarge the BCDW phase
for $X/t<0$ even in the $U,V\rightarrow 0$ limit, without disturbing the
Y- and the I-shaped structure of the phase diagram for the charge and
the spin parts (see Fig.~\ref{fig:g-ology}). Because this term makes the
magnitude of the backward and the Umklapp scattering couplings different
[$g_{1\perp}=U-2V+4X/\pi, g_{3\perp}=-(U-2V-4X/\pi)$] conserving the
SU(2)$\otimes$SU(2) symmetry of the Hubbard model.\cite{Japaridze-K} As
was predicted by the g-ology, the BCDW phase appears from the weak- to
the intermediate-coupling region.\cite{Aligia_comment} On the other
hand, in the strong-coupling region, the two transition lines are
coupled, and the direct CDW-SDW transition takes place.  The
strong-coupling theory also agrees with the Gaussian line as is shown in
Fig.~\ref{fig:diffs}(a).  This is the same behavior as in the case of
the pure EHM.  Therefore, the tricritical point in the pure EHM is
considered to have the same property of the one that separates the CDW,
the SDW, and the BCDW phases in the case of $X/t<0$.

For $X/t>0$, the order of the Gaussian and the spin-gap transitions
becomes vice versa, so that the BSDW phase appears.  However, in
Fig.~\ref{fig:diffs}(b), we find that the two transition lines cross in
the intermediate-coupling region, and a finite BCDW phase appears. This
BCDW region becomes narrower as the system size is increased.  In the
present analysis, we cannot conclude whether this BCDW phase remains or
vanishes in the thermodynamic limit. We will consider this phenomenon
again in Sec.~\ref{sec:SUMMARY}.

\subsection{Phase separation}

\begin{figure}
\noindent
\epsfxsize=3.3in \leavevmode \epsfbox{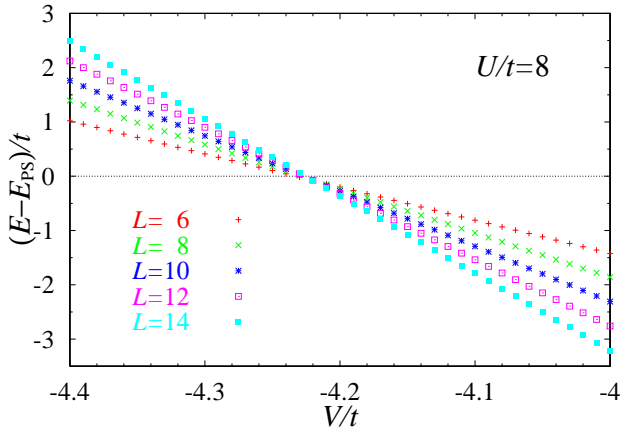}
\caption{Ground-state energy of the EHM for $L=6$-$14$ systems at
 $U/t=8$ measured from the fully phase separated state $E_{\rm
 PS}=(U+4V)L/2$ [Eq.~(\protect{\ref{eqn:E_PS}})]. The zero point gives
 the critical point for the phase separation.}
 \label{fig:phsep}
\end{figure}

Here, we determine the phase-separation boundary from the numerical data
of the exact diagonalization.\cite{Lin-H_comment} Usually, the phase
boundary is determined by the divergence of the
compressibility.\cite{Ogata-L-S-A} However, for the $U/t\gg1$ region of
the EHM, the phase separation can take place in the SDW state where the
compressibility cannot be defined. Consequently, the method of observing
the divergence of the compressibility is no longer valid. In this paper,
we determine the phase boundary by comparing the energy of the ground
state and that of the phase-separated state.  In the phase-separated
state at half-filling, the system is separated into doubly occupied
sites and a vacuum.  In this case, the energy in the thermodynamic limit
is exactly obtained as
\begin{equation}
  E_{\rm PS}=\frac{U+4V}{2}L.\label{eqn:E_PS}
\end{equation}
Therefore, we use the relation $E=E_{\rm PS}$ as a criterion for the
phase separation.\cite{Hirata-N} We show in Fig.~\ref{fig:phsep} the
ground-state energy of the EHM measured from the fully phase separated
state. The zero point gives the critical point for the phase separation.
The finite-size effect of the phase boundary is sufficiently small.

\begin{figure}[t]
\noindent
\epsfxsize=3.3in \leavevmode \epsfbox{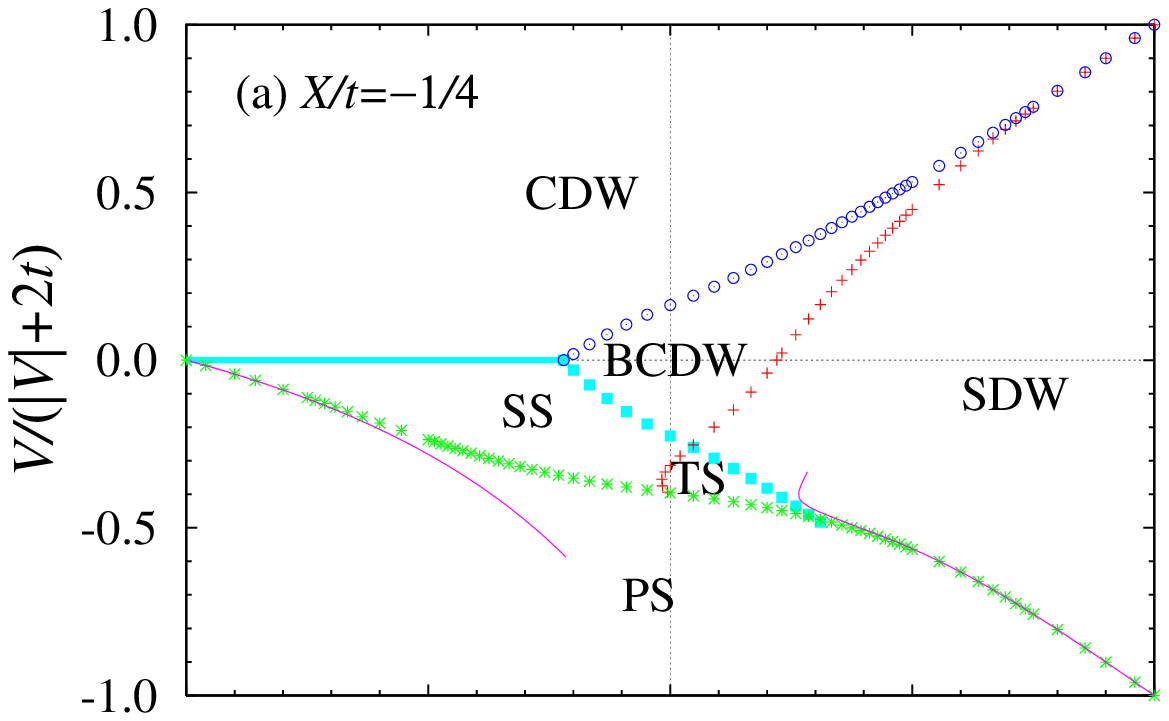}\\
\epsfxsize=3.3in \leavevmode \epsfbox{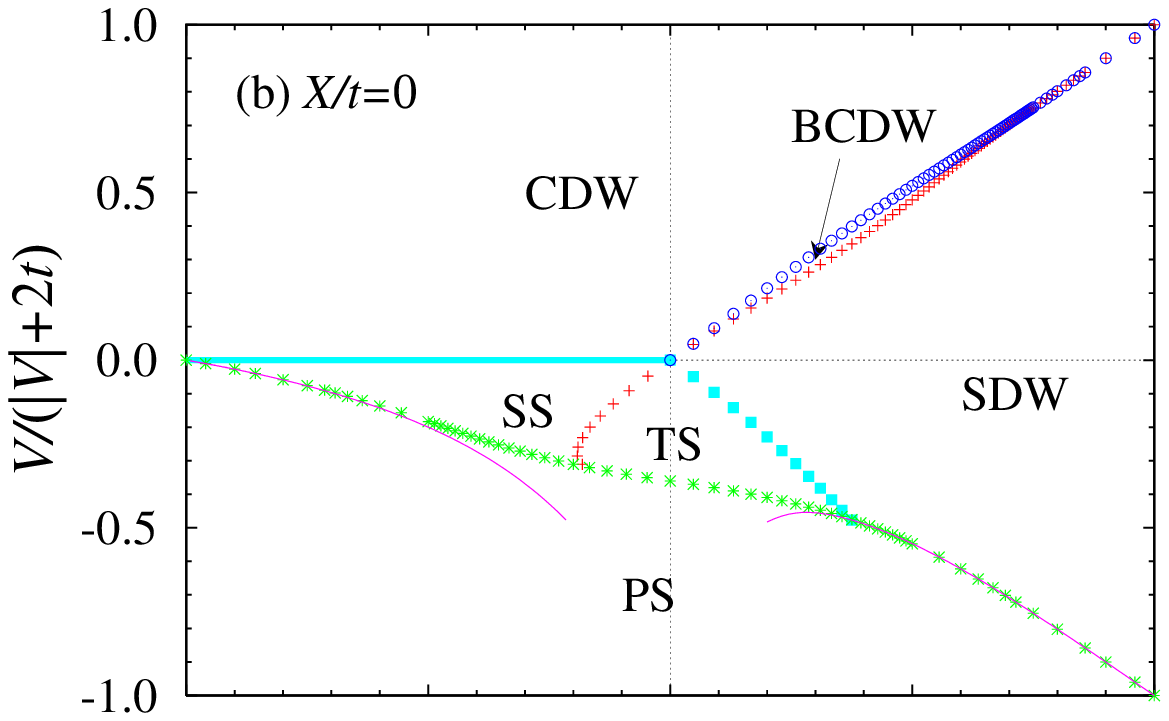}\\
\epsfxsize=3.3in \leavevmode \epsfbox{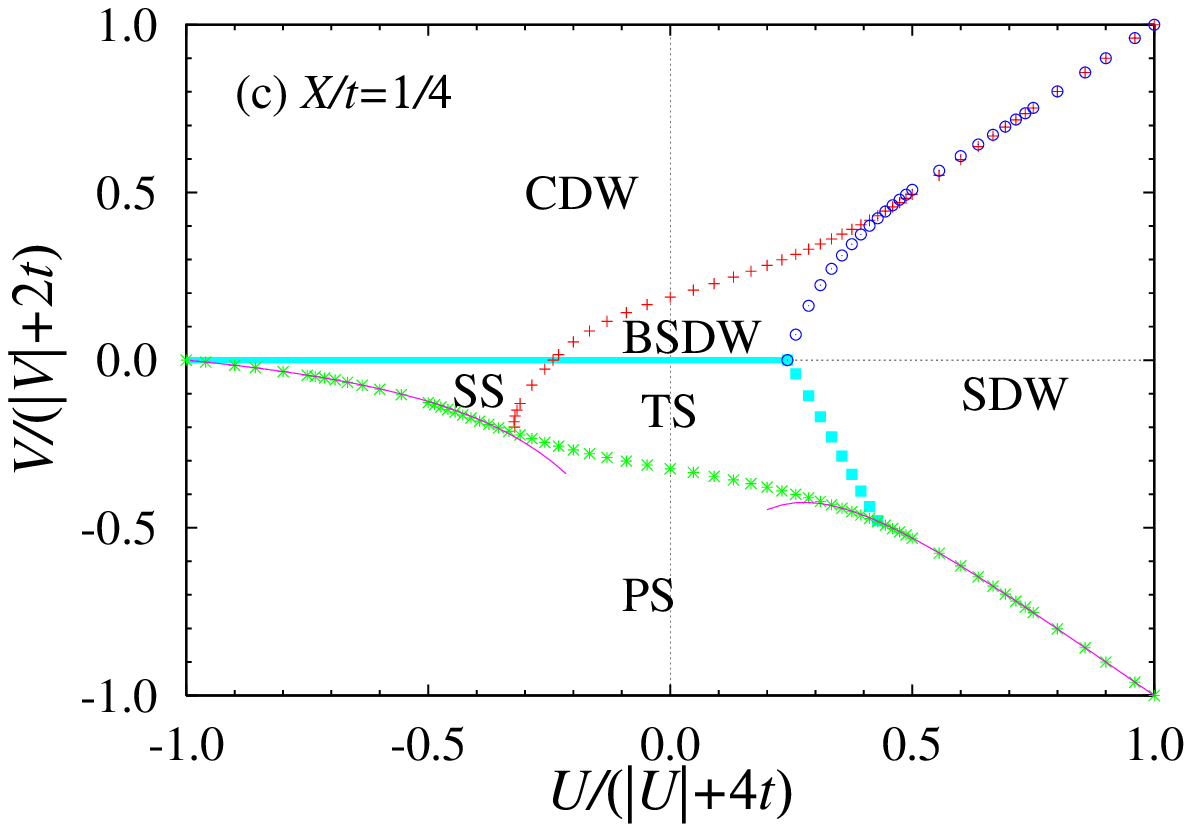}
\caption{Phase diagram of the 1D EHM determined by the data of the
$L=12$ system at half-filling for (a) $X/t=-1/4$, (b) $X/t=0$, and (c)
$X/t=1/4$ [CDW (SDW), charge (spin)-density wave; BCDW (BSDW),
bond-charge (spin)-density wave; SS (TS), singlet (triplet)
superconducting phase; PS, phase-separated state]. The asymptotic phase
boundaries for the PS are given by Eqs.~(\protect{\ref{eqn:ps2}}) and
(\protect{\ref{eqn:ps1}}).}  \label{fig:PD1}
\end{figure}

In order to check the validity of the numerical results, we compare our
result with the asymptotic phase boundary in the strong-coupling limit.
For $U/t\gg 1$ region, using Eqs.~(\ref{eqn:E_SDW}) and (\ref{eqn:E_PS}),
the phase boundary is given by\cite{Lin-H}
\begin{equation}
 E_{\rm PS}=E_{\rm SDW}.\label{eqn:ps2}
\end{equation}
On the other hand, for $U/t\ll-1$ region, the system can be mapped onto
the $S=1/2$ XXZ spin chain with antiferromagnetic coupling
$J_{xy}=4t^2/|U|$ and $J_z=J_{xy}+4V$, by using the $\eta$-pairing
operators (\ref{eqn:eta-pairing}) and the second-order perturbation
theory.  Then, the phase boundary is given by $J_{xy}=-J_z$, which
corresponds to the first-order transition between the XY and the
ferromagnetic phases in the spin system.\cite{Emery76,Fowler} If the
effect of the correlated hopping term (\ref{eqn:X-term}) is included,
the phase boundary is given by
\begin{equation}
 V=-\frac{2t^2}{|U|}(1-\xi)^2.\label{eqn:ps1}
\end{equation}
The numerical result given by $E=E_{\rm PS}$ well agrees with these
asymptotic phase boundaries (see Fig.~\ref{fig:PD1}).

\section{QUARTER-FILLING}\label{sec:QUARTER}

\begin{figure}
\noindent
\epsfxsize=3.3in \leavevmode \epsfbox{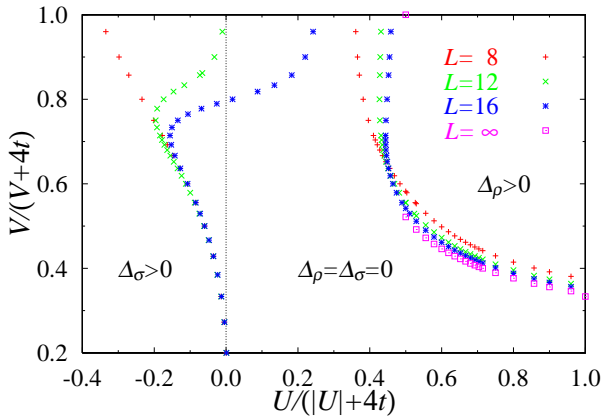}
\caption{Phase diagram of the 1D EHM at quarter-filling for the $V/t>0$
 region. The phase boundary between the TL liquid and the $4k_{\rm
 F}$-CDW phases is determined by the level crossing of $x_{\rho
 0}=4x_{\rho 2}$ in $L=8,12,16$ systems.  The critical points in the
 strong-coupling limits are $V_{\rm c}/t=2$ and $U_{\rm c}/t=4$,
 respectively.}  \label{fig:PD2}
\end{figure}

\begin{figure}
\noindent
\epsfxsize=3.3in \leavevmode \epsfbox{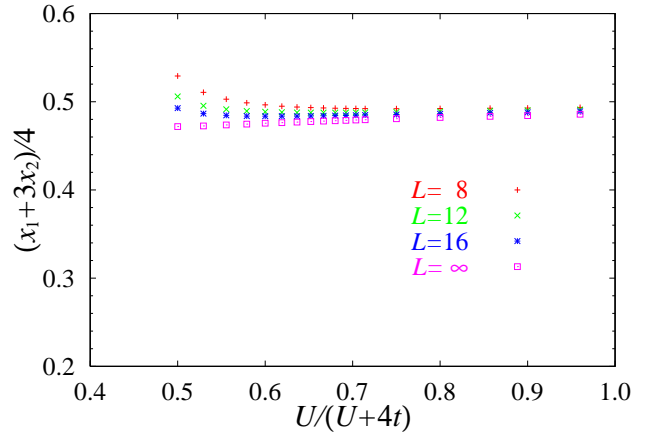}
\caption{Scaling dimensions given by
 Eq.~(\protect{\ref{eqn:check_BKTqf}}) on the BKT line.  The extrapolated
 value becomes $1/2$ with error less than $4\%$.}
 \label{fig:CHECK_qf}
\end{figure}

Finally, we analyze the phase diagram of the EHM at
quarter-filling\cite{Mila-Z,Penc-M,Sano-O} by the level-crossing
approach. In this case, the metal-insulator transition is considered as
a BKT transition due to the higher-order Umklapp scattering
($q=2$). Then, the phase boundary should be given by the level crossing
between the marginal and four times of the $4k_{\rm F}$-CDW spectra
$x_{\rho 0}=4x_{\rho 1}$ or $4x_{\rho2}=x_{\rho3}$, as was discussed in
Sec.~\ref{sec:non-su2}.  In the present numerical analysis, we use the
former level crossing, because $x_{\rho3}$ needs larger Hilbert space
than $x_{\rho0}$ and $x_{\rho2}$.  Based on this assumption, we obtain
the result shown in Fig.~\ref{fig:PD2}.  Since the level-crossing point
for the transition is higher ($x_{\rho}=2$) than the case of the
half-filling ($x_{\rho}=1/2$), the finite-size effect from the
irrelevant field (the deviation from the linearized dispersion relation)
becomes larger. In this case, we need an extrapolation of the critical
point to make the phase diagram. In the $U/t\rightarrow\infty$ limit,
the transition point for the charge-gap phase is given by $V_{\rm
c}/t=2$, because it corresponds to the XY-N\'{e}el transition in the
$S=1/2$ XXZ spin chain.\cite{Ovchinnikov73,Luther-P} For the $U/t\gg 1$
region, the extrapolated phase boundary flows into the point
$(U,V)=(\infty,2t)$ as we expected.  In order to check the consistency
in the finite-$U$ region, we calculate the following averaged scaling
dimension
\begin{equation}
\frac{x_{\rho 1}+3x_{\rho 2}}{4}=\frac{1}{2}.\label{eqn:check_BKTqf}
\end{equation}
Except for the large-$V$ region, the extrapolated value becomes $1/2$
with error less than $4\%$, as shown in Fig.~\ref{fig:CHECK_qf}.  Thus,
the universality class of the transition is considered to be a BKT-type.

On the other hand, for $V/t\gg 1$ region, since the finite-size effect
is too large, it is hard to perform the systematic
extrapolation. However, the phase boundary appears to flow into the
exact transition point $U_{\rm c}= 4t$ in the $V/t\rightarrow\infty$
limit as the system size is increased. Now let us review how the
critical point of the charge-gap phase in the $V/t\rightarrow\infty$
limit is derived.\cite{Mila-Z} The charge gap is defined by
\begin{equation}
 \Delta_{\rho} = E(N+1) + E(N-1) - 2 E(N).
\end{equation}
At quarter-filling, $E(L/2)=0$. If one electron is add to this, then the
energy is $E(L/2+1)=U$. Conversely, if one electron is removed, two free
holes appear, then they have a kinetic energy $E(L/2-1)\sim -
4t\cos(\pi/L)$. Therefore, the critical point for the charge-gap phase
is given by $U_{\rm c}= 4t$ in the thermodynamic limit.  We should note
that the critical point in the $U/t\rightarrow\infty$ and
$V/t\rightarrow\infty$ limits are given by a completely different
argument. Therefore, we expect that a tricritical behavior may also be
seen along the BKT transition line.

In a similar way as the half-filling, we consider the effect of a
charge-spin coupling operator which is derived from the $q=2$ Umklapp
scattering between three parallel spins and one antiparallel spin [the
$g_{3\parallel}$ term in Eq.~(\ref{eqn:eff_Ham})].  This operator may
cause the synchronization of the BKT transition ($q=2$) in the charge
part and the spin-gap transition. In order to examine this possibility,
we determine the spin-gap phase boundary by the singlet-triplet level
crossing in the large-$V$ region. Then, the phase boundary of the spin
gap appears to be coupled with that of the charge gap, and flows into
the point $(U,V)=(4t,\infty)$, as the system size is
increased. Therefore, this result suggests that the crossover in the
metal-insulator transition also takes place by the mechanism similar to
the case of half-filling. In the large-$V$ region, there are
phase-separated states,\cite{Clay-S-C} but that has not been studied in
the present analysis.  We should also consider the effect of the phase
separation in the future.

\section{SUMMARY AND DISCUSSION}\label{sec:SUMMARY}



Let us summarize the results obtained in this paper. We have determined
the phase diagrams of the 1D EHM at half- and quarter-filling using the
level-crossing approach, which is based on the TL liquid theory and the
renormalization group. The metal-insulator transitions in half-
(quarter-) filling are classified as BKT-type transitions due to the
first- (second-) order Umklapp scattering. This fact reflects the
``fractional quantization'' discussed by the bosonization theory
\cite{Giamarchi,Kolomeisky-S,Oshikawa-Y-A,Yamanaka-O-A} and the
generalized Lieb-Schultz-Mattis theorem.\cite{Oshikawa-Y-A,Yamanaka-O-A}


In the case of half-filling, for the charge sector, there are two BKT
lines reflecting the SU(2) and hidden SU(2) symmetries, and one Gaussian
line.  These three critical lines meet at the multicritical point
$(U/t,V/t)=(0,0)$, and they form a Y-shaped structure.  Note that the
same structure is also known in the phase diagram of the $S=1/2$
frustrated XXZ spin chain.\cite{Haldane82,Nomura-O} For the spin sector,
a spin-gap transition occurs due to the attractive backward scattering.
Since the transition takes place at the origin of the RG flow diagram,
the phase boundary has an I-shaped structure. Thus, the entire phase
diagram is given by the combination of the ``Y'' and the ``I''.


The transition between the CDW and the SDW phases has been considered by
assuming the following two scenarios: one is the independent Gaussian
and spin-gap transitions under the charge-spin separation, the other is
a direct CDW-SDW transition under the charge-spin coupling. By checking
the relations between the scaling dimensions, and by comparing the
numerical results with the strong-coupling theory, we have concluded
that the first scenario is realized from the weak- to the
intermediate-coupling ($U\sim 4t$) region, and the second scenario is
realized in the strong-coupling region. In the former case, a BCDW phase
exists between the CDW and the SDW phases.  Thus, the crossover of the
CDW-SDW transitions from the second order to the first order turned out
to be a phenomenon that reflects the validity of the charge-spin
separation.

To clarify the mechanism of the crossover in more detail, we have
investigated the phase diagram of the EHM including the correlated
hopping term (\ref{eqn:X-term}). In this case, there appear BCDW and
BSDW phases, depending on the order of the Gaussian and the spin-gap
transitions. Note that the direct CDW-SDW transition also takes place in
the strong-coupling region.

Therefore, we can understand the crossover of the CDW-SDW transition in
the EHM as a kind of these generalized cases.  The reason why the
mechanism of the CDW-SDW transition has been left ambiguous for long
times is that the analytical solutions both in weak- and strong-coupling
limits give the same phase boundary $U=2V$, and the numerical analysis
for the intermediate-coupling region does not have enough precision to
distinguish the Gaussian and the spin-gap transition lines.  Recently,
the similar mechanism of transition was reported in studies of the
Hubbard chain with periodic potential, which has the transition between
Mott and band insulators.\cite{Fabrizio-G-N,Takada-K}


At quarter-filling, the phase boundary of the metal-insulator transition
has also been determined by assuming that it is the BKT-type transition
of the higher-order Umklapp scattering.  The obtained phase boundary is
consistent with the known exact results in the $U\rightarrow\infty$ and
the $V\rightarrow\infty$ limits. In the present parameter space, there
are neither a BKT transition with the hidden symmetry, nor a Gaussian
transition.  Although the finite-size effect in $V/t\gg 1$ region is
large, the phase boundaries of the charge- and the spin-gap phases
appear to be coupled as the system size is increased.  Therefore, there
may be a crossover from the BKT to the first-order transitions in this
region.


The rest of the section is devoted to discussions.  We have clarified
the mechanism of the CDW-SDW transition of the EHM at half-filling.
However, we have not revealed the reason why the intermediate state in
Figs.~\ref{fig:3lines} and \ref{fig:PD1}(b) is not a BSDW but a BCDW.
We can interpret the appearance of the BCDW phase considering roles of
the $g_{3\parallel}$ term in Eq.~(\ref{eqn:eff_Ham}).  When the charge
gap opens in the $U>2V$ region ($g^*_{3\perp}=-\infty$), the phase field
is locked as $\phi_{\rho}=2n\pi/\sqrt{8}$ with $n$ being an integer,
which minimize the classical potential energy associated with the
$g_{3\perp}$ term.  In this case, the charge part of the
$g_{3\parallel}$ term is also locked, so that $g_{1\perp}=U-2V$ is
reduced as
\begin{equation}
 g_{1\perp}\rightarrow g_{1\perp}+g_{3\parallel}
  \langle \cos[\sqrt{8}\phi_{\rho}]\rangle,\label{eqn:shift1}
\end{equation}
where $g_{3\parallel}=-2V$.  Thus, if we estimate the spin-gap phase
boundary in the weak-coupling region using Eq.~(\ref{eqn:shift1}), it
shifts toward the $U>2V$ side of the $U=2V$ line. Similarly, we can
estimate the shift of the Gaussian line. In the $U<2V$ region, the spin
gap opens ($g^*_{1\perp}=-\infty$), then $g_{3\perp}=-U+2V$ shifts as
\begin{equation}
 g_{3\perp}\rightarrow g_{3\perp}+g_{3\parallel}
  \langle \cos[\sqrt{8}\phi_{\sigma}]\rangle,\label{eqn:shift2}
\end{equation}
where the phase field is locked as $\phi_{\sigma}=2n\pi/\sqrt{8}$.
Therefore, the Gaussian line shifts toward the opposite side of the
spin-gap phase boundary.  Consequently, the BCDW phase appears between
the CDW and the SDW phases.  Thus, it turns out that the
$g_{3\parallel}$ term enhances the BCDW phase when it is irrelevant, and
it couples the Gaussian line and the spin-gap phase boundary when it is
relevant.  The deviations from the $U=2V$ line may be analyzed
quantitatively in the weak-coupling region by the renormalization group
analysis including $g_{3\parallel}$.  The above explanation may also be
applicable to the phenomenon that a BCDW phase appears for $X/t>0$ in
the strong-coupling region [see Fig.~\ref{fig:diffs}(b)].


We should consider the reason why the size dependence of the Gaussian
transition line is smaller than that of the spin-gap phase boundary, in
the strong-coupling region of Figs.~\ref{fig:3lines} and
\ref{fig:diffs}.  The reason is considered to be the difference between
the behavior of the charge and the spin gaps.  As was discussed in
Sec.~\ref{sec:theory}, the spin gap opens exponentially slow
(\ref{eqn:spin-gap}), while the valley of the charge gap near the
Gaussian transition becomes steeper as the strength of the interaction
is increased (\ref{eqn:gaussian_trans}). In the present analysis based
on the TL liquid theory, we have assumed that both charge and spin parts
are gapless, so that the Gaussian (spin-gap) transition line may be
affected by the spin (charge) gap in the strong-coupling region.  In the
present case, magnitude and variation of the charge gap are considered
to be much larger than those of the spin gap near the $U=2V$ line in the
strong-coupling region, so that the Gaussian line has less size
dependence than the spin-gap phase boundary.


In the present paper, we have not determined the tricritical point.
This problem has been discussed by many
authors.\cite{Hirsch,Fourcade-S,Cannon-F,Cannon-S-F,Voit92} Recently,
the tricritical point is explored by the
density-matrix-renormalization-group
method,\cite{GPZhang,GPZhang_comment} but we may also determine the
tricritical point by comparing the numerical result of the Gaussian
transition line and the strong-coupling perturbative expansions in the
higher order.  It may also be worth studying the level crossing for the
tricritical point by considering the logarithmic corrections to the
excitation spectra which stem from the $g_{3\parallel}$ term.  In
addition to the tricritical point between the CDW and the SDW phases,
there is another tricritical point that separates the TS, the SDW, and
the PS states. In Fig.~\ref{fig:PD1}, the phase boundaries between the
SDW and the PS states obtained by the strong-coupling theory seem to be
bent near the tricritical point. This tricritical point may also be
identified by the further strong-coupling calculation.


Finally, we discuss the effect of site-off-diagonal interactions. In
this paper, we have considered the correlated hopping term given by
Eq.~(\ref{eqn:X-term}), however the generalized form of this term is
given by\cite{Simon-A}
\begin{mathletters}
\begin{eqnarray}
{\cal H}_{X}&=&X\sum_{is}
(c^{\dag}_{is} c_{i+1,s} + \mbox{H.c.})(n_{i,-s} + n_{i+1,-s}),\\
{\cal H}_{X'}&=&X'\sum_{is}
(c^{\dag}_{is} c_{i+1,s} + \mbox{H.c.})n_{i,-s}n_{i+1,-s}.
\end{eqnarray}
\end{mathletters}
In the present paper, we have set $X=-X'/2$ to keep the particle-hole
symmetry and the SU(2)$\otimes$SU(2) symmetry of the Hubbard model.  If
this relation is not chosen, these symmetries are lost, so that the
$V=0$ line is no longer the phase boundary of the metal-insulator
transition at half-filling.  Besides, an additional Umklapp term
$\sin\sqrt{8}\phi_{\rho}$ appears in the effective
Hamiltonian.\cite{Voit92,Japaridze-K} The analysis for this situation
($X\neq -X'/2$) is the subject of future research.

There are other types of site-off-diagonal interactions. For example,
the bond-bond interaction term is given by\cite{Campbell-G-L}
\begin{equation}
 {\cal H}_{W}=W\sum_{iss'}
 (c^{\dag}_{is} c_{i+1,s}+\mbox{H.c.})
 (c^{\dag}_{is'} c_{i+1,s'}+\mbox{H.c.}).
 \label{eqn:W-term}
\end{equation}
This term also makes the difference in the magnitude of the g-parameters
for the backward and the Umklapp scattering in the weak-coupling limit
[$g_{1\perp}=U-2V+8W, g_{3\perp}=-(U-2V-8W)$](Ref.~\ref{Voit92})
conserving the symmetries of the Hubbard model. Therefore, we expect
that this term play a role similar to the correlated hopping terms
[Eq.~(\ref{eqn:X-term})].  On the other hand, since this term can be
rewritten as the exchange of the spins and of the pseudospins
(\ref{eqn:eta-pairing}), it may affect the first-order transition to a
ferromagnetic state\cite{Campbell-G-L} or the phase separation.
Besides, in a parameter region of the EHM including this term, the BSDW
state is shown to be the exact ground state.\cite{Itoh-N-M} The analysis
of the effect of this term by the level-crossing approach will be
reported elsewhere.\cite{Nakamura_99b}

\section{ACKNOWLEDGMENTS}

The author is grateful to S.~Hirata, K.~Itoh, T.~Kawarabayashi,
A.~Kitazawa, K.~Kusakabe, N.~Muramoto, H.~Nakano, K.~Okamoto,
M.~Oshikawa, H.~Otsuka, H.~Shiba, M.~Shiroishi, M.~Takahashi, and
J.~Voit for useful discussions.  The computation in this work was partly
done with the facilities of the Supercomputer Center, Institute for
Solid State Physics, University of Tokyo.

\appendix

\section{Weak-coupling limit}\label{sec:weak-coupling}
In the weak-coupling limit, the parameters of the sine-Gordon model
(\ref{eqn:eff_Ham}) can be identified in terms of the bare coupling
constants of the original model, and consequently the phase boundaries
are obtained analytically.\cite{Emery,Solyom,Voit,Voit92,Japaridze-K}
This approach is often called the g-ology.

The parameters of Eq.~(\ref{eqn:eff_Ham}) are given by the g-parameters
defined in Refs.~\ref{Solyom}, \ref{Voit}, and \ref{Voit92} as follows
\begin{eqnarray}
 v_{\nu}&=&\sqrt{u_{\nu}^2-\left(\frac{g_{\nu}}{2\pi}\right)^2},\ \
 K_{\nu}=\sqrt{\frac{2\pi u_{\nu}+g_{\nu}}{2\pi u_{\nu}-g_{\nu}}},
 \nonumber\\
 u_{\nu}&\equiv&v_{\rm F}+\frac{g_{4\parallel}\pm g_{4\perp}}{2\pi},\ \
 g_{\nu}\equiv g_{1\parallel}-g_{2\parallel}\mp g_{2\perp},
\end{eqnarray}
where the upper (lower) sign corresponds to $\nu=\rho$ ($\nu=\sigma$),
and $v_{\rm F}=2t\sin(k_{\rm F})$ is the Fermi velocity.  For the EHM
including the $X$ [Eq.~(\ref{eqn:X-term})] and $W$
[Eq.~(\ref{eqn:W-term})] terms at half-filling, the g-parameters can be
identified as follows:
\begin{eqnarray}
 g_{1\perp}=g_{\sigma}&=&U-2V+\delta g,\label{eqn:g_sigma}\nonumber\\
 g_{3\perp}&=&-(U-2V-\delta g),\label{eqn:g3a}\nonumber\\
 g_{\rho}&=&-(U+6V-\delta g),\label{eqn:g_rho}\label{eqn:g3b}\\
 \delta g&=&4X/\pi+8W.\nonumber
\end{eqnarray}
This calculation can be performed straightforwardly except for the $X$
term. For the $X$ term which contains a three-body term, the
operator-product-expansion technique is needed to identify the
g-parameters.\cite{Japaridze-K}

The instabilities for the charge and the spin gaps are discussed based
on the renormalization group (RG) analysis as explained in
Sec.~\ref{sec:theory}.  Since $K_{\nu}$ is approximated as
$K_{\nu}\approx 1+g_{\nu}/2\pi v_{\nu}$, the parameter in the RG flow
diagram (Fig.~\ref{fig:RGflow}) is given by $y_{0\nu}(l)=g_{\nu}/\pi
v_{\nu}$.  Then, the spin-gap opens when $ g_{1\perp}=g_{\sigma}<0$, and
the phase boundary is
\begin{equation}
U=2V-\delta g.
\end{equation}
On the other hand, the charge gap opens when $g_{3\perp}>|g_{\rho}|$.
The condition $g_{3\perp}=-g_{\rho}<0$ is the BKT-type transition due to
the SU(2) symmetry in the charge sector.  Then, the phase boundary is
obtained as
\begin{equation}
V=0,\ U<\delta g.
\end{equation}
The condition $g_{3\perp}=g_{\rho}>0$ is the BKT-type transition due to
the hidden SU(2) symmetry in the charge sector. Then, we obtain the
phase boundary as
\begin{equation}
U=-2V+\delta g,\ U>\delta g.  
\end{equation}
The Gaussian-type transition takes place at $g_{3\perp}=0$, for
$g_{\rho}<0$.  Thus, we obtain the Gaussian line for the charge sector
as
\begin{equation}
U=2V+\delta g,\ V<0.
\end{equation}
Thus, we obtain the phase diagrams in the weak-coupling limit as is
shown in Fig.~\ref{fig:g-ology} which have the Y- and the I-shaped
structures in the charge and the spin degrees of freedom, respectively.
In the present g-ology analysis, the parameters $X$ and $W$ appear only
through $\delta g$, so that the $X$ and $W$ terms play similar roles in
respect of the enhancement of the BCDW ($\delta g <0$) or the BSDW
($\delta g >0$) phases.  The correspondence between the g-ology and the
level-crossing approach is summarized in Table \ref{tbl:gology_vs_lc}.

In this paper, we have taken $U_{r,s}$ into account in the
 identification of $g_{1\perp}$ and $g_{3\perp}$ by following
 Ref.~\ref{Senechal}.  The Clifford algebra for $U_{r,s}$
 (\ref{eqn:ladder_op}) can be expressed by tensor products of the Pauli
 matrices, which are chosen so the effective Hamiltonian
 (\ref{eqn:eff_Ham}) is to be diagonal in the space of $U_{r,s}$.  For
 example, the following representation is possible:
\begin{equation}
 \begin{array}{ccc}
 U_{{\rm R}\uparrow}=\tau^x\otimes\tau^x, &~~~~~~&
 U_{{\rm R}\downarrow}=\tau^z\otimes\tau^x,\\
 U_{{\rm L}\uparrow}=\tau^y\otimes\tau^x, & &
 U_{{\rm L}\downarrow}=1\otimes\tau^y.
 \label{eqn:ladder1}
 \end{array}
\end{equation}
These operators contribute to the $g_{1\perp}$ and the $g_{3\perp}$
terms as $\pm 1\otimes\tau^z$, and one eigenvalue of the matrix is
chosen.  Consequently, the signs of $g_{1\perp}$ and $g_{3\perp}$ become
opposite as in Eqs.~(\ref{eqn:g3b}).  When we consider the bosonization
of a physical operator, it should be diagonalized simultaneously with
the Hamiltonian. Therefore, if the physical operator cannot be
diagonalized by Eq.~(\ref{eqn:ladder1}), we should choose other
representations for $U_{r,s}$. In the derivation of the physical
operators in Sec.~\ref{sec:PHASES}, we need one more representation such
as
\begin{equation}
 \begin{array}{ccc}
 U_{{\rm R}\uparrow}=\tau^z\otimes\tau^x, &~~~~~~&
 U_{{\rm R}\downarrow}=\tau^x\otimes\tau^x,\\
 U_{{\rm L}\uparrow}=\tau^y\otimes\tau^x, & &
 U_{{\rm L}\downarrow}=1\otimes\tau^y.
 \label{eqn:ladder2}
 \end{array}
\end{equation}
In this case, the contribution to the $g_{1\perp}$ and the $g_{3\perp}$
terms is $\mp 1\otimes\tau^z$. If the contribution of $U_{r,s}$ is
neglected, the sign of the $g_{3\perp}$ term is reversed, and the roles
of ${\cal O}_{\rho 1}$ and ${\cal O}_{\rho 2}$ are interchanged.


\begin{references}
\bibitem{BOW_comment}
 In the preceding works (Refs.~\ref{Voit92} and \ref{Nakamura_99}), BCDW
 is called ``bond-order wave (BOW)''.
\bibitem{Dongen_comment}
We solved the equation numerically without using approximations.  An
approximation used in Ref.~\ref{Dongen} is not appropriate for the
quantitative analysis, as the author himself discussed.  In the
approximated case, the region where the CDW-SDW transition agrees with
the Gaussian line is overestimated up to $U/t\sim 6$.
\bibitem{Aligia_comment}
In Ref.~\ref{Aligia-H-B-O}, the authors determined the phase diagram of
the 1D Hubbard model ($V=0$) including the correlated hopping term at
half-filling [SU(2)$\otimes$SU(2) symmetric case], by observing the
quantum numbers corresponding to the current excitations ($m_{\nu}$)
using the Berry phase. Therefore, if the SU(2) symmetry in the charge
sector is broken ($V\neq 0$), their method may correspond to the level
crossing of the Gaussian-type transition, but may not correspond to the
one of the BKT-type transition.
\bibitem{GPZhang_comment}
 In Ref.~\ref{GPZhang}, the phase boundary between the CDW and the SDW
 phases is determined by observing the CDW and the SDW structure factors
 using the density-matrix-renormalization-group method. This result
 agrees with our level-crossing point of the CDW and the SDW
 excitations.
\end{references}
\end{document}